%% file: B2G-18-008_temp.tex
\pdfoutput=1

\documentclass[11pt,twoside,a4paper,cmspaper,final,collab]{cms-tdr}
\def\svnVersion{f6c5828-D}\def\svnDate{2019/09/29}\def\cmsCernNoTag{CERN-EP-2019-056}\def\cmsCernDate{\today}\def\cmsMessage{"Published in the Journal of High Energy Physics as \href{http://dx.doi.org/10.1007/JHEP10(2019)125}{\doi{10.1007/JHEP10(2019)125}.}"}

\begin{document}\cmsNoteHeader{B2G-18-008}

\hyphenation{had-ron-i-za-tion}
\hyphenation{cal-or-i-me-ter}
\hyphenation{de-vices}
\RCS$HeadURL$
\RCS$Id$

\newlength\cmsFigWidth
\ifthenelse{\boolean{cms@external}}{\setlength\cmsFigWidth{0.49\textwidth}}{\setlength\cmsFigWidth{0.65\textwidth}}
\ifthenelse{\boolean{cms@external}}{\providecommand{\cmsLeft}{upper\xspace}}{\providecommand{\cmsLeft}{left\xspace}}
\ifthenelse{\boolean{cms@external}}{\providecommand{\cmsRight}{lower\xspace}}{\providecommand{\cmsRight}{right\xspace}}
\ifthenelse{\boolean{cms@external}}{\providecommand{\NA}{\ensuremath{\cdots}\xspace}}{\providecommand{\NA}{\ensuremath{\text{---}}\xspace}}

\newcommand{\mx}{\ensuremath{m_{\PX}}\xspace}

\newcommand{\hbb}{\ensuremath{\PH\to\bbbar}\xspace}
\newcommand{\hww}{\ensuremath{\PH\to\PW\PW^*}\xspace}
\newcommand{\wqq}{\ensuremath{\PW\to\qqbar'}\xspace}

\newcommand{\hbbjet}{\ensuremath{\bbbar~\text{jet}}\xspace}
\newcommand{\hbbjets}{\ensuremath{\bbbar~\text{jets}}\xspace}
\newcommand{\wqqjet}{\ensuremath{\qqbar'~\text{jet}}\xspace}
\newcommand{\wqqjets}{\ensuremath{\qqbar'~\text{jets}}\xspace}
\newcommand{\wlnucand}{\ensuremath{\ell\nu~\text{candidate}}\xspace}
\newcommand{\hwwcand}{\ensuremath{\PW\PW^* ~\text{candidate}}\xspace}

\newcommand{\wjets}{\ensuremath{\PW\! +\! \text{jets}}\xspace}
\newcommand{\zll}{\ensuremath{\PZ\to\ell\ell}\xspace}

\newcommand{\tauTwooOne}{\ensuremath{\tau_{2}/\tau_{1}}\xspace}
\newcommand{\wqqtau}{\ensuremath{\qqbar'~\tau_{2}/\tau_{1}}\xspace}
\newcommand{\md}{\ensuremath{m_{\text{D}}}\xspace}
\newcommand{\ptom}{\ensuremath{\pt/m}\xspace}
\newcommand{\pz}{\ensuremath{p_z}\xspace}

\newcommand{\mhbb}{\ensuremath{m_{\bbbar}}\xspace}
\newcommand{\mhh}{\ensuremath{m_{\PH\PH}}\xspace}
\newcommand{\mhiggs}{\ensuremath{m_{\PH}}\xspace}

\newcommand{\mtn}{\ensuremath{m_{\cPqt}}\xspace}
\newcommand{\mwn}{\ensuremath{m_{\PW}}\xspace}
\newcommand{\losttwn}{\ensuremath{\text{lost}~\cPqt/\PW}\xspace}
\newcommand{\Losttwn}{\ensuremath{\text{Lost}~\cPqt/\PW}\xspace}
\newcommand{\qgn}{\ensuremath{\Pq/\Pg}\xspace}

\newcommand{\bL}{\ensuremath{\cPqb\text{L}}\xspace}
\newcommand{\bM}{\ensuremath{\cPqb\text{M}}\xspace}
\newcommand{\bT}{\ensuremath{\cPqb\text{T}}\xspace}

\newcommand{\nonqg}{\ensuremath{\text{non-}\qgn}\xspace}
\newcommand{\Nonqg}{\ensuremath{\text{Non-}\qgn}\xspace}

\newcommand{\mtbkg}{\ensuremath{\mtn~\text{background}}\xspace}
\newcommand{\mwbkg}{\ensuremath{\mwn~\text{background}}\xspace}
\newcommand{\losttwbkg}{\ensuremath{\losttwn~\text{background}}\xspace}
\newcommand{\qgbkg}{\ensuremath{\qgn~\text{background}}\xspace}

\newlength\cmsTabSkip\setlength{\cmsTabSkip}{1ex}

\cmsNoteHeader{B2G-18-008}

\title{Search for resonances decaying to a pair of Higgs bosons in the $\bbbar\qqbar'\ell\nu$ final state in proton-proton collisions at $\sqrt{s}=13\TeV$}

\date{\today}

\abstract{
A search for new massive particles decaying into a pair of Higgs bosons in proton-proton collisions at a center-of-mass energy of 13\TeV is presented. Data were collected with the CMS detector at the LHC, corresponding to an integrated luminosity of 35.9\fbinv. The search is performed for resonances with a mass between 0.8 and 3.5\TeV using events in which one Higgs boson decays into a bottom quark pair and the other decays into two $\PW$~bosons that subsequently decay into a lepton, a neutrino, and a quark pair. The Higgs boson decays are reconstructed with techniques that identify final state quarks as substructure within boosted jets. The data are consistent with standard model expectations. Exclusion limits are placed on the product of the cross section and branching fraction for generic spin-0 and spin-2 massive resonances. The results are interpreted in the context of radion and bulk graviton production in models with a warped extra spatial dimension. These are the best results to date from searches for an $\PH\PH$ resonance decaying to this final state, and they are comparable to the results from searches in other channels for resonances with masses below 1.5\TeV.
}

\hypersetup{
pdfauthor={CMS Collaboration},
pdftitle={Search for resonances decaying to a pair of Higgs bosons in the bbqq'lnu final state in proton-proton collisions at sqrt(s)=13 TeV},
pdfsubject={CMS},
pdfkeywords={CMS, physics, BSM, Beyond the standard model, graviton, radion, diboson, HH, diHiggs, di-Higgs, WWbb, boosted}}

\maketitle

\section{Introduction\label{sec:intro}}
The discovery of a Higgs boson (\PH)~\cite{HiggsDiscoveryAtlas,HiggsDiscoveryCMS,Chatrchyan:2013lba} established the existence of at least a simple mass generation mechanism for the standard model (SM)~\cite{Englert:1964et,Higgs}, the so-called ``Higgs Mechanism.'' 
The simple model, however, has a number of limitations that are ameliorated\cite{Branco:2011iw}
by a so-called ``extended Higgs sector.''
Supersymmetry~\cite{Ramond:1971gb,Golfand:1971iw,Neveu:1971rx,Volkov:1972jx,Wess:1973kz,Wess:1974tw,Fayet:1974pd,Nilles:1983ge} requires such an extended Higgs sector, with new spin-0 particles.   Another class of models with warped extra dimensions, proposed by Randall and Sundrum~\cite{Randall:1999ee}, postulates the existence of a compact fourth spatial dimension with a warped metric. Such compactification creates
heavy resonances arising as a tower of Kaluza--Klein excitations, leading to possible spin-0 radions~\cite{Goldberger:1999uk,DeWolfe:1999cp,Csaki:1999mp,Csaki:2000zn} or spin-2 bulk gravitons~\cite{Davoudiasl:1999jd,Agashe:2007zd,Fitzpatrick:2007qr}.
The ATLAS~\cite{ATLASwprimeWZPAS,Aad:2014yja,Aad:2015xja,Aad:2015yza,Aad:2015uka,Aaboud:2017ahz,Aaboud:2017fgj,Aaboud:2017gsl,Aaboud:2017itg,Aaboud:2017rel,Aaboud:2017cxo,Aaboud:2017eta,Aaboud:2018ohp,Aaboud:2018sfw,Aaboud:2018knk,Aaboud:2018zhh} and CMS~\cite{Khachatryan:2014xja,Khachatryan:2015ywa,Khachatryan:2015year,Khachatryan:2015bma,Khachatryan:2015tha,Khachatryan:2016yji,Khachatryan:2016cfx,Sirunyan:2016cao,Sirunyan:2017wto,Sirunyan:2017nrt,Sirunyan:2017acf,Sirunyan:2017isc,Sirunyan:2017jtu,Sirunyan:2018iff,Sirunyan:2018qob,Sirunyan:2018ivv,Sirunyan:2018hsl,Sirunyan:2018fuh,Sirunyan:2018qca} Collaborations have conducted a number of searches for these particles, where the new bosons decay into vector bosons
and/or Higgs bosons ($\PW\PW$, $\PZ\PZ$, $\PW\PZ$, $\PH\PH$, $\PZ\PH$, or $\PW\PH$).

In this paper, we describe a search for narrow resonances ($\PX$) decaying to $\PH\PH$, where one $\PH$ decays to a bottom quark pair ($\bbbar$) and the other decays to a $\PW$ boson pair, with at least one $\PW$ boson off-shell ($\PW\PW^*$). These are the most likely and second-most likely Higgs boson decay channels, respectively.
The otherwise large SM background of jets produced via quantum chromodynamics processes, referred to as ``multijet'' background, is greatly reduced by considering the $\PW\PW^*$ final state in which one $\PW$ boson decays to quarks ($\cPq\cPaq'$) and the other to either an electron-neutrino pair ($\Pe\Pgn$) or a muon-neutrino pair ($\Pgm\Pgn$). This search is optimized for particle mass $\mx>0.8\TeV$  and employs new techniques for this channel to recognize substructure within boosted jets. The search is performed on a data set collected in 2016 at the CERN LHC, corresponding to an integrated luminosity of $35.9\fbinv $ of proton-proton ($\Pp\Pp$) collisions at $\sqrt{s}=13\TeV$.

The Higgs bosons have a high Lorentz boost because of the large values of $\mx$ considered, and the decay products of each one are produced in a collimated cone. The $\hbb$ decay is reconstructed as a single jet, referred to as the \hbbjet, with high transverse momentum \pt. The \hww decay is also reconstructed as a single jet, referred to as the \wqqjet, but with a nearby lepton ($\Pe$ or $\Pgm$). In both cases, the jets are required to have a reconstructed topology consistent with a substructure arising from a boson decaying to two quarks. The semileptonic Higgs boson decay chain is reconstructed from both the visible decay products and the missing transverse momentum. A distinguishing characteristic of the signal is a peak in the two-dimensional plane of the \hbbjet mass \mhbb and the reconstructed $\PH\PH$ invariant mass \mhh.

The main SM background to this search arises from top quark pair \ttbar production in which one top quark decays via a charged lepton ($\cPqt\to \PW\cPqb \to \ell\Pgn\cPqb$) and the other decays exclusively to quarks ($\cPqt \to \PW\cPqb \to \cPq\cPaq'\cPqb$).
The top quarks affecting this analysis have decay products that are collimated because of large boosts.  In particular, the all-hadronic top quark decays can be misreconstructed as single  $\hbbjet$s.
Peaks in the \mhbb distribution from this background correspond to fully contained top quark and $\PW$ boson decays. The second-largest background  is primarily composed of production of $\PW$ bosons in association with jets (\wjets) and multijet events.  Both \wjets and multijet background events are experimentally distinct from \ttbar production, in part because their \mhbb distributions are smoothly falling.

In this analysis, the events are divided into 12 exclusive categories by lepton flavor, \wqqjet substructure, and \hbbjet flavor identification. The SM background and signal yields are then simultaneously estimated using a maximum likelihood fit to the two-dimensional distribution in the \mhbb and \mhh mass plane.

\section{The CMS detector}
The central feature of the CMS apparatus is a superconducting solenoid of 6\unit{m} internal diameter, providing a magnetic field of 3.8\unit{T}. Within the solenoid volume are a silicon pixel and strip tracker, a lead tungstate crystal electromagnetic calorimeter (ECAL), and a brass and scintillator hadron calorimeter (HCAL), each composed of a barrel and two endcap sections. Forward calorimeters extend the coverage in pseudorapidity $\eta$ provided by the barrel and endcap detectors. Muons are detected in gas-ionization chambers embedded in the steel flux-return yoke outside the solenoid. Events of interest are selected using a two-tiered trigger system~\cite{Khachatryan:2016bia}. The first level, composed of custom hardware processors, uses information from the calorimeters and muon detectors to select events at a rate of around 100\unit{kHz} within a time interval of less than 4\mus. The second level, known as the high-level trigger, consists of a farm of processors running a version of the full event reconstruction software optimized for fast processing, and reduces the event rate to around 1\unit{kHz} before data storage. A more detailed description of the CMS detector, together with a definition of the coordinate system used and the relevant kinematic variables, can be found in Ref.~\cite{Chatrchyan:2008zzk}.

\section{Simulated samples\label{sec:samples}}
Signal and background yields are extracted from data with a fit using templates of the two-dimensional \mhbb and \mhh mass distribution. The signal and background templates are obtained from samples generated using Monte Carlo simulation.

\tolerance=800
The signal process $\Pp\Pp\to\PX\to\PH\PH\to\bbbar\PW\PW^{*}$ is simulated for both the spin-0 and spin-2 resonance scenarios. The $\PX$ bosons are produced via gluon fusion and have a $1\MeV$ resonance width, which is small compared to the experimental resolution. The samples are generated at leading order (LO) using the \MGvATNLO 2.2.2 generator~\cite{Alwall:2014hca} with MLM merging~\cite{Alwall:2007fs} for \mx between 0.8 and 3.5\TeV.
\par

The background processes are produced with a variety of generators.
The same generator as for signal is used to produce \ttbar, \wjets, multijet, Higgs boson production in association with a $\PQt$ quark ($\cPqt\PH$), and Drell--Yan samples. Samples of $\PW\PZ$ diboson production and the associated production of \ttbar with either a $\PW$ or $\PZ$ boson ($\ttbar+\text{V}$) are also generated with \MGvATNLO, but at next-to-leading-order (NLO) with the FxFx jet merging scheme~\cite{Frederix2012}. The $\PW\PW$~diboson process, single top production, and $\ttbar\PH$ are generated with $\POWHEG~\cmsSymbolFace{v2}$ at NLO~\cite{Nason:2004rx,Frixione:2007vw,Alioli:2010xd,Re:2010bp,Melia:2011tj,Nason:2013ydw,Frederix:2012dh,Hartanto:2015uka}.
Single top in the associated production ($\cPqt\PW$) and $t$-channel ($\cPqt\cPq$) processes are included, but not $s$-channel ($\cPqt\cPqb$), which is negligible.  

For all samples, the parton showering and hadronization are simulated with $\PYTHIA~\cmsSymbolFace{8.205}$~\cite{Sjostrand:2014zea} using the CUETP8M1~\cite{Khachatryan:2015pea} tune, with NNPDF~3.0~\cite{Ball:2011mu} parton distribution functions (PDFs). The simulation of the CMS detector is performed with the $\GEANTfour$~\cite{Agostinelli:2002hh} toolkit. Additional $\Pp\Pp$ collisions in the same or nearby bunch crossings (pileup) are simulated and the samples are weighted to have the same pileup multiplicity as measured in data.

While the final background normalizations are extracted from data with the template fit, all 
processes are initially normalized to their theoretical cross sections, 
using the highest order available.
The \ttbar process is rescaled to the next-to-next-to-leading-order (NNLO) cross section, computed with $\textsc{Top++}~\textsc{v2.0}$~\cite{Czakon:2011xx}. The \wjets and Drell--Yan samples are also normalized using NNLO cross sections, but calculated with $\FEWZ~\textsc{v3.1}$~\cite{Li:2012wna}. NLO cross sections are used for the single top and diboson samples, calculated with $\MCFM~\textsc{v6.6}.$~\cite{Campbell:2011bn,Campbell:2012uf,Campbell:2004ch}. The multijet and $\ttbar+\text{V}$ cross sections are obtained from \MGvATNLO at LO and NLO accuracy, respectively. NLO cross sections are used for the  $\ttbar\PH$ and $\cPqt\PH$ processes~\cite{deFlorian:2016spz}.

\section{Event reconstruction\label{sec:reco}}
Signal events and those from the primary SM background source, \ttbar production with a single-lepton final state,
have similar signatures.  Both processes feature high-\pt jets with substructure consistent with two or more quarks, jets containing {\cPqb} hadron decays, and leptons that originate from a \PW~boson decay.  Additional discrimination of signal events from background events is achieved by associating the lepton and each jet with a particle in the $\PH\PH$ $\to$ $\bbbar\PW\PW^* \to \bbbar\ell\nu\cPq\cPaq'$ decay chain and applying mass constraints.

A particle-flow (PF) algorithm~\cite{CMS-PRF-14-001} aims to reconstruct and identify each individual particle in an event, with an optimized combination of information from the various elements of the CMS detector. The reconstructed vertex with the largest value of summed tracking-object $\pt^2$ is taken to be the primary $\Pp\Pp$ interaction vertex. These tracking objects are track jets and the negative vector sum of the track jet \pt. Track jets are clustered using the anti-\kt jet finding algorithm~\cite{Cacciari:2008gp,Cacciari:2011ma} with the tracks assigned to the vertex as inputs.

\subsection{Electron and muon identification}

Events are required to have exactly one isolated lepton. This lepton is associated with the leptonic $\PW$ boson decay.
Reconstructed electrons are required to have $\pt>20\GeV$ and $\abs{\eta}<2.5$, and are identified with a high-purity selection to suppress the potentially large multijet background~\cite{Khachatryan:2015hwa}. Muons are required to have $\pt>20\GeV$ and $\abs{\eta}<2.4$, and to pass identification criteria optimized to select muons with $>$95\% efficiency~\cite{Sirunyan:2018}.
The impact parameter of lepton tracks with respect to the primary vertex is required to be consistent with originating from that vertex: longitudinal distance ${<}0.1\cm$, transverse distance ${<}0.05\cm$, and significance ${<}4$ standard deviations of the three-dimensional displacement.  These criteria remove background events where the lepton is produced by a semileptonic heavy-flavor decay rather than a $\PW$ boson decay.  In addition, these criteria prevent incorrectly selecting a lepton from a heavy-flavor decay in signal events. Requiring leptons to be isolated from nearby hadronic activity is important to suppress background, but can also cause significant signal inefficiency because of the collinear decay of the Lorentz-boosted Higgs boson. This inefficiency is mitigated by using an isolation definition specifically designed for leptons from boosted decays~\cite{Rehermann:2010vq}. The isolation metric $I_{\text{rel}}$ is the \pt sum of the PF particles with $\Delta R<\Delta R_\text{iso}$ with respect to the lepton, divided by the lepton \pt. The angular distance is $\Delta R=\sqrt{\smash[b]{(\Delta\eta)^2+(\Delta\phi)^2}}$. The value $\Delta R_\text{iso}$ is defined to be
\begin{linenomath}
\begin{equation}
  \Delta R_\text{iso} =
    \begin{cases}
      0.2, & \pt < 50 \GeV, \\
      10\GeV/\pt, & 50< \pt < 200 \GeV, \\
      0.05, & \pt>200\GeV,\\
    \end{cases}
\end{equation}
\end{linenomath}
which preserves signal efficiency even in the case of high \mx. The neutral particle contribution to $I_{\text{rel}}$ from pileup interactions is estimated and removed using the method described in Ref.~\cite{Khachatryan:2015hwa}. Electrons are selected with $I_{\text{rel}}<0.1$, whereas muons, because of lower background rates, are selected with $I_{\text{rel}}<0.2$.

Muons in signal events have an approximate efficiency of $85\%$ for $\mx=0.8\TeV$, decreasing to $70\%$ for $\mx=3.5\TeV$, with isolation being the leading source of inefficiency compared to all other requirements. The efficiency to select electrons is lower, approximately $40\%$ for $\mx=0.8\TeV$, decreasing to $6\%$ for $\mx=3.5\TeV$. The leading source of electron inefficiency is a selection imposed at the reconstruction level on the ratio of the energy deposited in the HCAL to that deposited in the ECAL. Signal electrons typically fail this selection because of the nearby energy deposits from the hadronic $\PW$ boson decay. Lepton reconstruction, identification, and isolation efficiencies are measured in a \zll data sample with a ``tag-and-probe'' method~\cite{Khachatryan:2010xn} and the simulation is corrected for any discrepancies with the data. There is generally much less hadronic activity in \zll events than in signal events, so these corrections are parameterized by nearby hadronic activity to ensure their applicability. For this measurement, a lepton's hadronic activity is quantified by using the PF particles with $\Delta R<0.4$ about the lepton to obtain two variables: the relative \pt sum around the lepton and the $\Delta R$ between the lepton and the $\vec p$ sum of these particles.  When parameterized by these two variables, a similar drop in efficiency is measured in low $\Delta R$ and high relative momentum \zll events as in signal events. The lepton selection efficiencies in simulation are found to be within 10\% of those in data. The uncertainty in the correction is at its largest for high hadronic activity, with a maximum value of 10\% for electrons and 5\% for muons.

\subsection{Jet clustering and momentum corrections}

Two types of jets are used. Because the \PX bosons being considered here are much more massive compared to the mass of the Higgs bosons they decay into, the subsequent \hbb and \wqq decays are each reconstructed as single, merged jets. These jets are formed by clustering PF particles according to the anti-\kt algorithm~\cite{Cacciari:2008gp,Cacciari:2011ma} with a distance parameter of 0.8, and are referred to as AK8 jets. The PF particle or particles associated with the lepton are not included in the clustering of this jet type in order to prevent the \wqqjet from containing the lepton's momentum. Jets of the second type, referred to as AK4 jets, are used to suppress background events from \ttbar production by identifying additional jets originating from $\PQb$ quarks. These jets are also clustered according to the anti-\kt algorithm, but with a distance parameter of 0.4. Jets of both types are required to have $\abs{\eta}<2.4$ so that a majority of their area is within acceptance of the tracker. The AK8 jets are required to have $\pt>50\GeV$, whereas the threshold is $20\GeV$ for AK4 jets.

Jet momentum for both jet types is determined as the vectorial sum of all particle momenta in the jet, and is found from simulation to be, on average, within 5 to 10\% of the true momentum over the whole \pt spectrum and detector acceptance. Additional $\Pp\Pp$ interactions within the same or nearby bunch crossings can contribute additional tracks and calorimetric energy depositions, increasing the apparent jet momentum. The pileup per particle identification (PUPPI) algorithm~\cite{Bertolini:2014bba} is used to mitigate the effect of pileup at the reconstructed particle-level, making use of local shape information, event pileup properties, and tracking information. Charged particles identified to be originating from pileup vertices are discarded. For each neutral particle, a local shape variable is computed using the surrounding charged particles compatible with the primary vertex within the tracker acceptance ($\abs{\eta} < 2.5$), and using both charged and neutral particles in the region outside of the tracker coverage. The momenta of the neutral particles are then rescaled according to their probability to originate from the primary interaction vertex deduced from the local shape variable~\cite{CMS-PAS-JME-16-003}. Jet energy corrections are derived from simulation studies so that the average measured response of jets becomes identical to that of particle level jets. In situ measurements of the momentum balance in dijet, photon+jet, $\PZ\text{+jet}$, and multijet events are used to determine any residual differences between the jet energy scale in data and in simulation, and appropriate corrections are made~\cite{Khachatryan:2016kdb}. Additional selection criteria are applied to each jet to remove jets potentially dominated by instrumental effects or reconstruction failures~\cite{CMS-PAS-JME-16-003}.

\subsection{Hadronic boson decay reconstruction}

In high-$\mx$ signal events, the \hww decay is reconstructed as an AK8 jet and a nearby lepton, with the jet itself containing two localized energy deposits, ``subjets,'' one from each quark.
Only the AK8 jet closest in $\Delta R$ to the lepton is considered for \wqqjet reconstruction.
This jet satisfies \wqqjet reconstruction criteria if it is close to the lepton ($\Delta R<1.2$) and if two subjets with $\pt>20\GeV$ and $\abs{\eta}<2.4$ can be identified. The constituents of the jet are first reclustered using the Cambridge--Aachen algorithm~\cite{Dokshitzer:1997in,Wobisch:1998wt}. The ``modified mass drop tagger'' algorithm~\cite{Dasgupta:2013ihk,Butterworth:2008iy}, also known as the ``soft drop'' (SD) algorithm, with angular exponent $\beta = 0$, soft cutoff threshold $z_{\mathrm{cut}} < 0.1$, and characteristic radius $R_{0} = 0.8$~\cite{Larkoski:2014wba}, is applied to remove soft, wide-angle radiation from the jet. The subjets used in the analysis are those remaining after the algorithm has removed all recognized soft radiation. The purity of the \wqqjet reconstruction is quantified using the ``$N$-subjettiness'' variables $\tau_{N}$, which measure compatibility with the hypothesis that a jet originates from $N$ subjets~\cite{Thaler:2010tr}. The $\tau_{N}$ are obtained by first reclustering the jet into $N$ subjets using the \kt algorithm~\cite{Catani:1993hr}.
The variables are then calculated with these subjets as described in Ref.~\cite{Thaler:2010tr} with a characteristic radius $R_{0} = 0.8$. The ratio of $N$-subjettiness variables, \tauTwooOne, is used to discriminate \wqqjets originating from two-pronged $\PW$ boson decays against those from single quarks or gluons.

Generally, the Higgs bosons in signal events have large Lorentz boosts and are produced with \mbox{$\Delta\phi\approx\pi$} between them.
Therefore, \hbbjet candidates are required to be AK8 jets with $\Delta\phi>2$ from the lepton and $\Delta R>1.6$ from the \wqqjet.
If there are two or more \hbbjet candidates, the one leading in \pt is used. This jet is reconstructed as a \hbbjet if it is the leading or second-leading AK8 jet in \pt, has $\pt>200\GeV$, and if two constituent subjets with $\pt>20\GeV$ and $\abs{\eta}<2.4$ can be identified. The \hbbjet SD mass, which is the invariant mass of the two subjets, is used to obtain \mhbb. The mass grooming helps reject events for which the \hbbjet originates from a single quark or gluon. The performance of the SD algorithm varies with \hbbjet \pt, so simulation-derived \mhbb correction factors are applied as a function of \pt to make the average \mhbb value be $125\GeV$, the Higgs boson mass $m_\PH$~\cite{Aad:2015zhl}.

\subsection{Jet flavor identification}

Jets and subjets are identified as likely to have originated from {\cPqb} hadron decays using the combined secondary vertex {\cPqb} tagging algorithm~\cite{Sirunyan:2017ezt}. Two operating points of the algorithm are used, which have similar performance on subjets and AK4 jets. A high-efficiency working point, referred to as ``loose,'' has an efficiency of ${\approx}80\%$ and a light-quark or gluon misidentification rate of ${\approx}10\%$. The ``medium'' operating point has an efficiency and misidentification rate of ${\approx}60\%$ and ${\approx}1\%$, respectively. A ``tight'' operating point is not used. Jets or subjets with $\pt>30\GeV$ and $\abs{\eta}<2.4$ are considered for {\cPqb} tagging. 
This lower bound on \pt is chosen because the uncertainty in b tagging calibrations is larger for lower \pt jets and because the b quarks in our signal events have large \pt.   
The {\cPqb} tagging efficiency and misidentification rate are measured in data, and the simulation is corrected for any discrepancy~\cite{Sirunyan:2017ezt}.

\subsection{Semileptonic Higgs boson decay and signal mass reconstruction}

The missing transverse momentum vector \ptvecmiss is computed as the negative vector \pt sum of all the PF candidates in an event~\cite{CMS-PAS-JME-17-001}. The \ptvecmiss is modified to account for corrections to the energy scale of the reconstructed jets in the event.
The \ptvecmiss is an estimate of the transverse momentum of the neutrino in the semileptonic Higgs boson decay chain. The longitudinal momentum \pz of this neutrino is estimated by setting the invariant mass of the neutrino, the lepton, and the \wqqjet to $m_\PH$ and solving the corresponding second-order equation. If two real solutions exist, the one with the smaller magnitude is chosen. If the \pz solution is complex, the real component of the solution is used. Other methods for determining the neutrino \pz, including choosing the other \pz solution or incorporating the imaginary components, do not improve the \mhh resolution. The reconstructed momentum of the $\PW$ boson that decays to leptons, referred to as the \wlnucand, is obtained from the lepton and the estimated neutrino momenta. The \hwwcand momentum is then obtained from the combined \wlnucand and the \wqqjet momenta. The invariant mass of this object and the \hbbjet is \mhh.

\section{Event selection and categorization\label{sec:selection}}
Events are included in this search if they pass the following criteria that indicate they originate from a $\PX$ boson decay and are then divided into 12 independent categories. A separate set of criteria is used to define control regions, which are used to validate the modeling of background processes.

\subsection{Event selection}

Events are selected by the trigger system if they contain one of the following: an isolated electron with $\pt>27\GeV$, an isolated muon with $\pt>24\GeV$, or $\HT>800\GeV$ ($900\GeV$ for the last quarter of data taking), where \HT is the scalar sum of jet \pt~for all AK4 online jets with $\pt>30\GeV$.
A combination (inclusive OR) of lepton and \HT triggers is used because the online lepton isolation selection is inefficient for high-\mx signal, which provides two high-\pt, collimated Higgs boson decays. These events have large \HT and are instead selected with higher efficiency by the \HT trigger.  Additional multi-object triggers that select events with a single lepton and $\HT>400\GeV$ supplement these two single-object triggers, thereby maintaining high signal trigger efficiency for the entire \mx analysis range. The pileup correction for  \HT is the same offline as in the trigger.
The trigger efficiency is measured for \ttbar events in data and is $>$94\% for events passing \HT and lepton $\pt$
offline selection criteria. The simulation is corrected so that its trigger efficiency matches the efficiency measured with data. The trigger efficiency for signal events is $98\%$ for $\mx=0.8\TeV$ and $>$99\% for $\mx>1\TeV$.

Offline, events are required to have $\HT>400\GeV$ and a lepton with $\pt>30\GeV$ for electrons and $\pt>26\GeV$ for muons. Background events from $\zll$ are suppressed by rejecting events that contain additional leptons with $\pt>20\GeV$. Events are further required to have a \wqqjet and a \hbbjet. Background from \ttbar production is reduced by vetoing events with AK4 jets that are $\Delta R>1.2$ from the \hbbjet and pass the medium {\cPqb} tagging operating point.

\begin{figure}[ht!]
\centering
\includegraphics[width=0.45\textwidth]{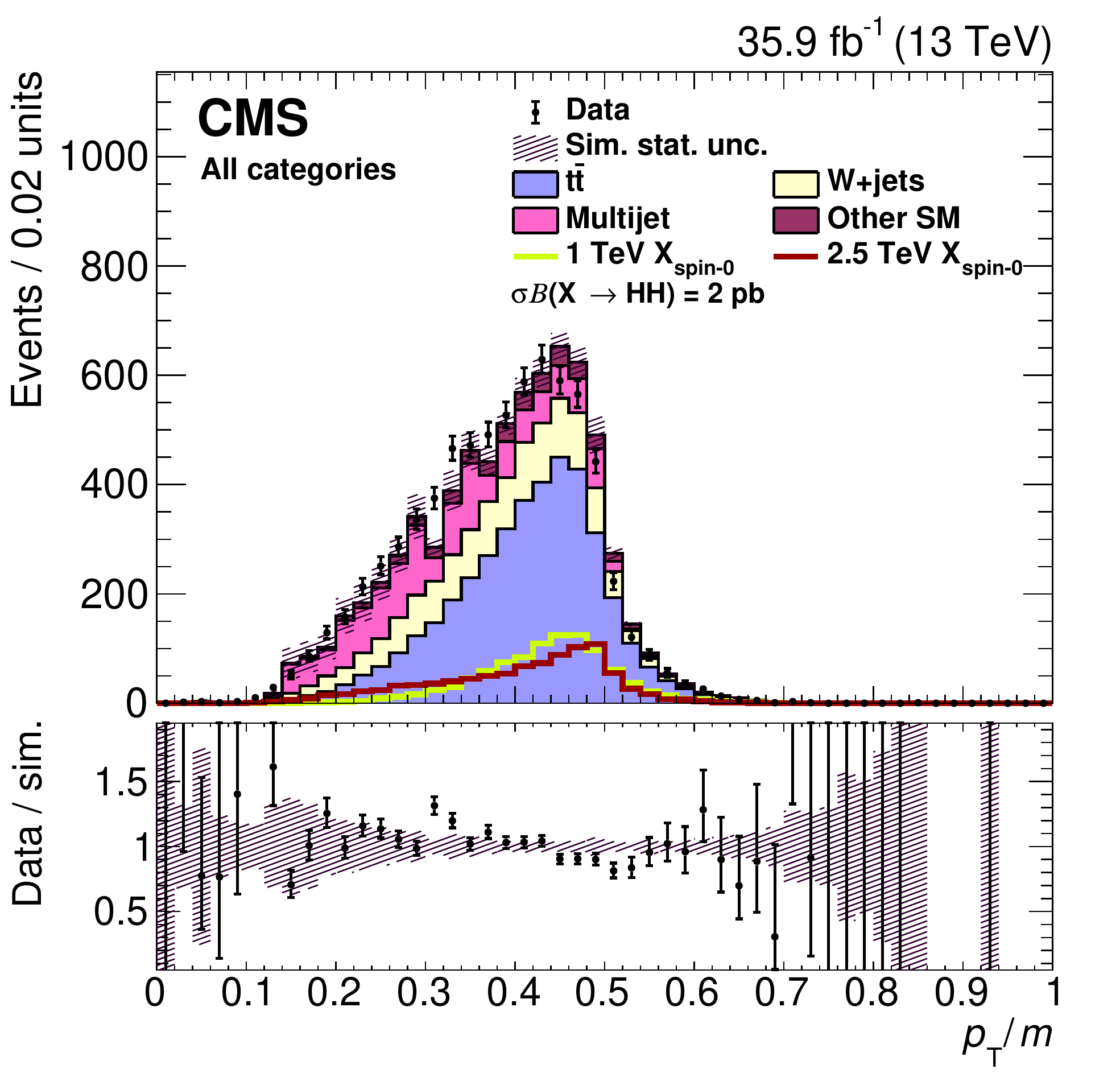}
\includegraphics[width=0.45\textwidth]{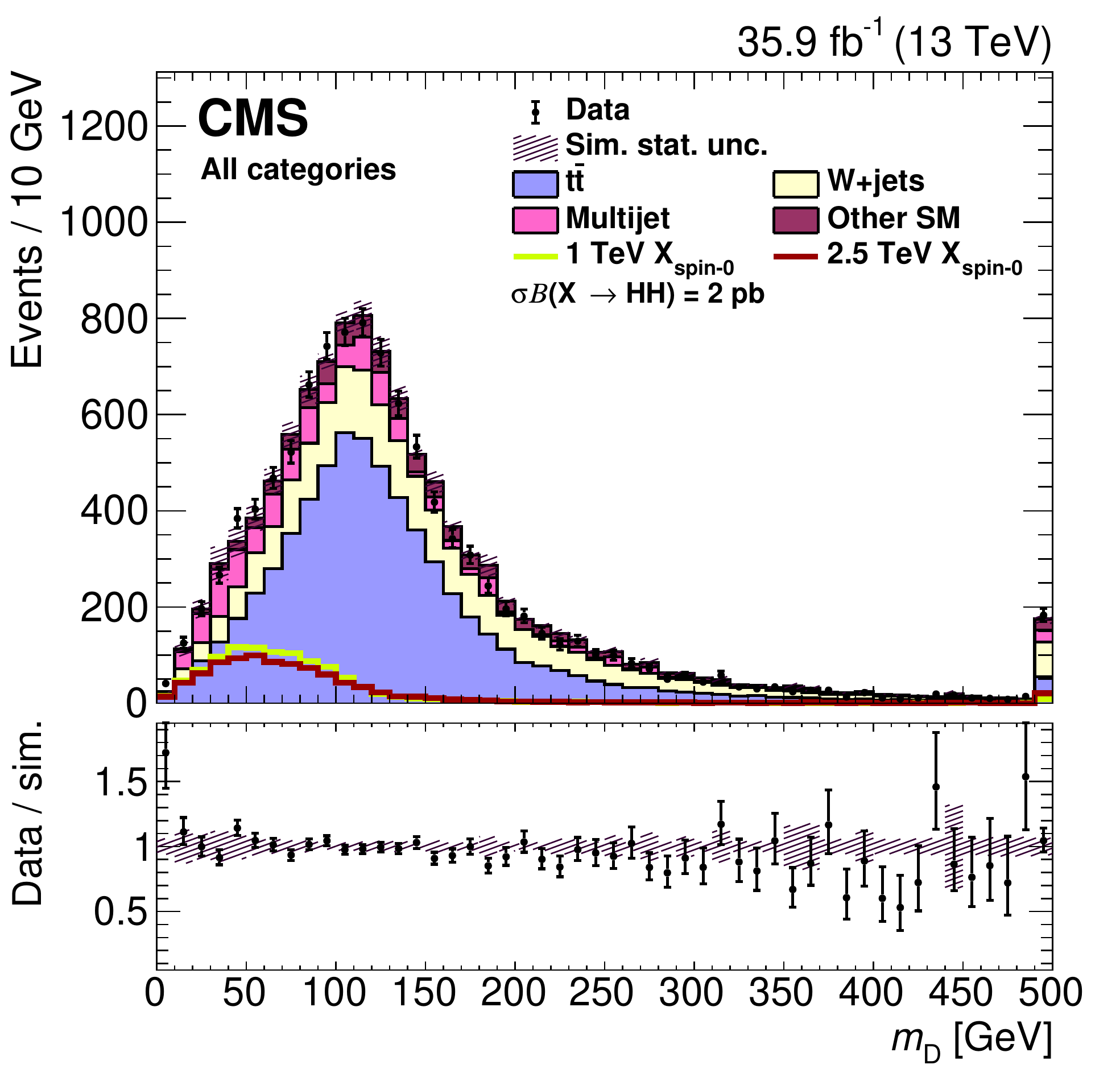}
\includegraphics[width=0.45\textwidth]{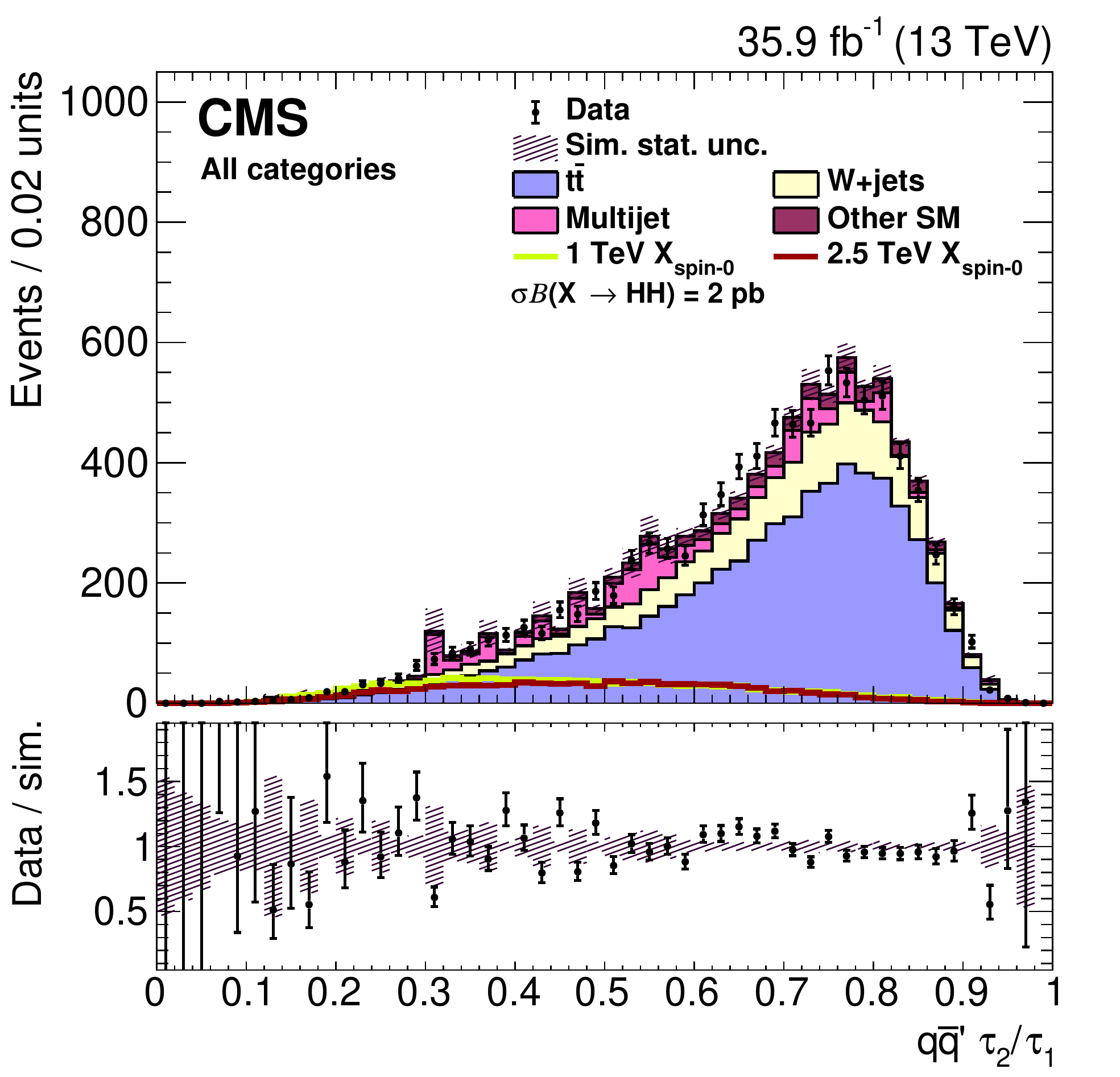}
\includegraphics[width=0.45\textwidth]{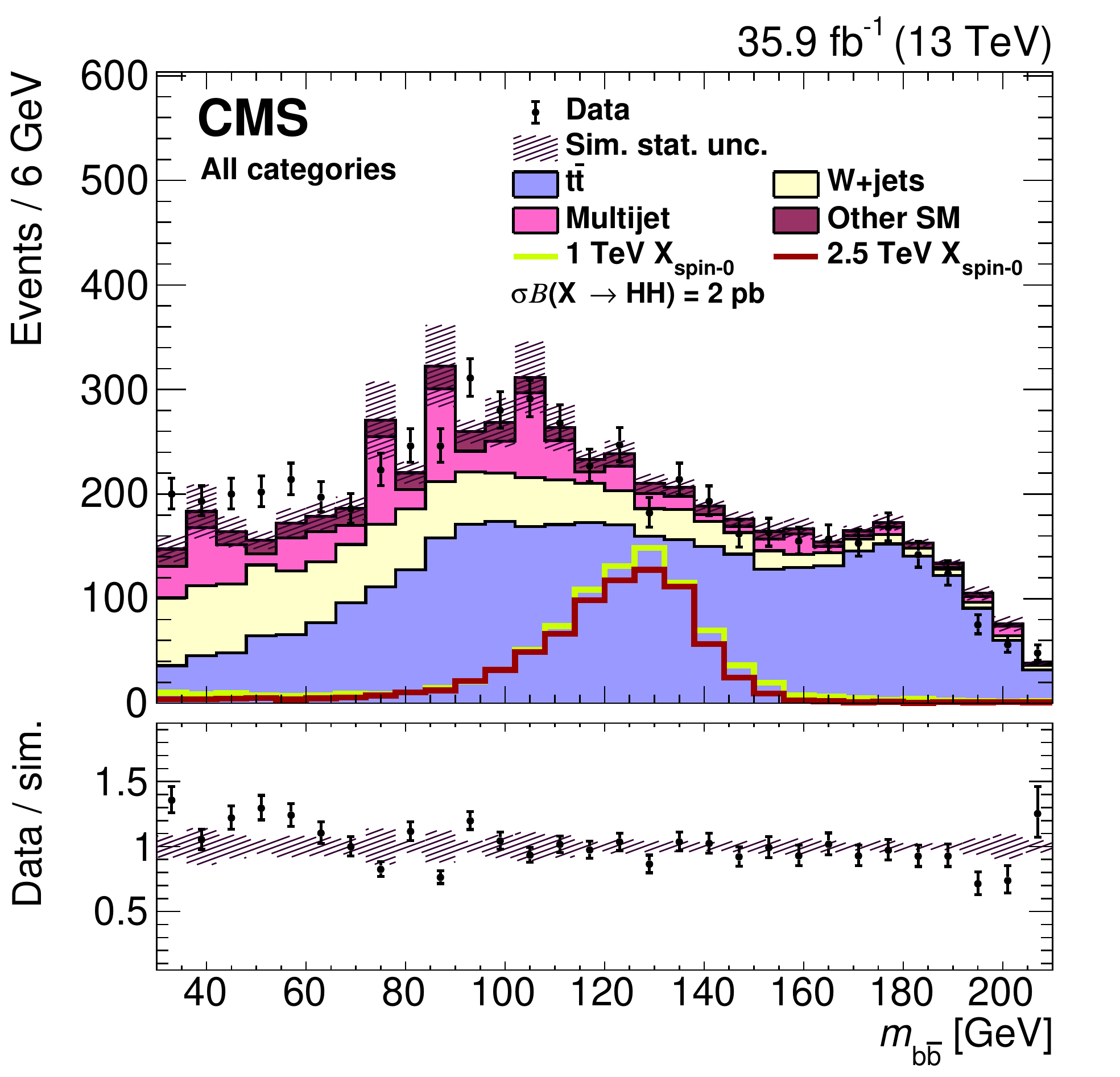}

\caption{Pre-modeling and pre-fit distributions of the discriminating variables, which are described in the text, are shown for data (points) and SM processes (filled histograms) as predicted directly from simulation. The statistical uncertainty of the simulated sample is shown as the hatched band.  The solid lines correspond to spin-0 signals for $\mx$ of 1 and 2.5\TeV. The product of the cross section and branching fraction to two Higgs bosons is set to 2\unit{pb} for both signal models. The lower panels show the ratio of the data to the sum of all background processes.}
\label{fig:event_vars}
\end{figure}

Jets in multijet and \wjets events tend to be produced at higher $\abs{\eta}$ than those produced in signal events, which contain jets from the decay of a heavy resonance. The ratio \ptom, which is the \hwwcand \pt divided by \mhh, exploits this property and is especially effective at high \mhh. Events are required to have $\ptom>0.3$. A \mhiggs constraint on the \hwwcand is not useful because it is already imposed in the neutrino momentum calculation. However, there is discrimination because the decay chain involves a two-body decay as an intermediate step. We define a variable $\md\equiv\pt\Delta R/2$, where $\Delta R$ is the separation of the two reconstructed $\PW$ bosons and the \pt is that of the \hwwcand. This variable is based on an approximate expression for the opening angle of a highly boosted, massive particle decay. The selection $\md<125\GeV$ is applied and has a high efficiency for signal events. The \md and \ptom distributions are shown in Fig.~\ref{fig:event_vars}. This figure is shown only to illustrate how these variables are used to discriminate signal events from background events; the simulated distributions are pre-modeling and pre-fit. The initial difference in $\mhbb$ near $50 \GeV$ between simulation and data is apparent only with the pre-fit background model; with the full post-fit background model no discrepancy appears.

\subsection{Event categorization}

Events are categorized by event properties that reflect the signal purity. The categorization allows for a single set of selections that targets the full \mx range, which is preferable to search categories that are optimized for different mass ranges. Electron and muon events are separated because
their efficiencies for background and signal are different, resulting in different signal purities. The electron and muon categories are labeled ``\Pe'' and ``$\mu$,'' respectively, in the figures. There are three categories of {\cPqb} tagging, evaluated by counting the number of subjets in the \hbbjet that pass {\cPqb} tagging operating points. The first, labeled ``\bL,'' is composed of events in which one subjet passes the medium operating point and the other does not pass the loose operating point. Events with one subjet passing the medium operating point and one passing the loose but not the medium operating point are denoted ``\bM,'' and those with two subjets passing the medium operating point are labeled ``\bT.'' The final categorization is based on the \tauTwooOne $N$-subjettiness ratio of the \wqqjet, referred to as \wqqtau. Events with $0.55 < \wqqtau < 0.75 $ fall into the low-purity category, ``LP,'' while those with $\wqqtau<0.55$ are included in a high-purity category, ``HP.'' The \wqqtau distribution is shown in Fig.~\ref{fig:event_vars}. Events are divided into all combinations of categories for a total of 12 exclusive selections. When describing a single selection, the category label is a combination of those listed above. For example, the tightest {\cPqb} tagging category with a low-purity \wqqtau selection in the electron channel is: ``\Pe,~\bT,~LP.'' The categories and their corresponding labels are summarized in Table~\ref{tab:cuts_cat}.

\begin{table}[!ht]
\centering
\topcaption{\label{tab:cuts_cat} Event categorization and corresponding category labels. All combinations of the two lepton flavor, three \hbbjet subjet {\cPqb} tagging, and two \wqqjet substructure selections are used to form 12 independent event categories. For the \hbbjet subjet {\cPqb} tagging type, ``medium'' refers to the subjets that pass the medium {\cPqb} tagging operating point and ``loose'' refers to those that pass the loose, but not the medium, operating point.}
\begin{tabular}{lll}
Categorization type & Selection  & Category label\\
\hline
Lepton flavor	& Electron & \Pe \\
 	 			& Muon & $\mu$ \\
\hbbjet subjet {\cPqb} tagging	& One medium & \bL \\
								& One medium and one loose & \bM \\
								& Two medium & \bT \\
\wqqjet substructure	& $0.55 < \wqqtau < 0.75 $ & LP \\
						& $\wqqtau<0.55$ & HP \\
\hline
\end{tabular}
\end{table}

The search is performed in these categories for $30 < \mhbb < 210\GeV$.  Events below $30\GeV$ would provide little sensitivity and would be relatively difficult to model since these are events for which the SD algorithm results in nearly all of the jet energy being removed. The \mhbb distribution is displayed in Fig.~\ref{fig:event_vars}. Events with $700 < \mhh < 4000\GeV$ are analyzed. The lower bound is chosen such that the \mhh distribution is monotonically decreasing for background events. The upper bound is far above the highest mass event observed in data. For spin-0 scenarios, the selection efficiency for $\PX\to\bbbar\qqbar'\ell\Pgn$ events to pass the criteria of any event category is $9\%$ at $\mx=0.8\TeV$. The efficiency increases with $\mx$ to $18\%$ at $\mx=1.2\TeV$ because the Higgs boson decays become more collimated. Above 1.2\TeV the selection efficiency decreases to a minimum of $9\%$ at $\mx=3.5\TeV$ because of the combination of lower {\cPqb} tagging efficiency for high-\pt jets and the worsening of the lepton isolation for extremely collimated Higgs boson decays. The Higgs bosons in spin-2 signal events are more central in polar angle than those from spin-0 signal, resulting in a larger selection efficiency, ${\approx}15\%$, relative.

\subsection{Control region event selection and categorization}

Two control regions are used to validate the SM background estimation and to obtain systematic uncertainties.
The first, labeled ``\ttbar~CR,'' targets backgrounds with top quarks, specifically \ttbar production. Such events are selected by inverting the AK4 jet \cPqb-tagging veto. The $\md$ and $\ptom$ selections are removed to increase the statistical power  of the sample. This control region is then divided into the 12 categories previously described. Overall, the \mhbb and \mhh shapes in this control region are very similar to the shapes in the signal region for the backgrounds that contain top quarks. The top quark \pt spectrum in \ttbar events has been shown to be mis-modelled in simulation~\cite{Sirunyan:2017mzl,Khachatryan:2016mnb}. A correction is measured in this region and applied to the simulation as a normalization correction. However, ultimately the final value of the normalization and its uncertainty come from the two-dimensional fit to signal and background. While the $\ttbar$~CR is an adequate probe of processes that involve top quarks, it is not sensitive to the multijet or \wjets backgrounds. 
Instead, a second control region, labeled ``\qgn~CR,'' is used to study the modeling of the mass shapes and the relative composition of the \wjets and the multijet backgrounds, which is similar to their relative composition in the search region.  
The selection of events in this control region is the same as for the signal region, except that the \hbbjet is required to have no subjets passing the loose {\cPqb} tagging operating point. As a result, the events in this control region are not categorized by \hbbjet~\cPqb~tagging, but are still categorized by lepton flavor and \wqqtau.

\section{Background and signal modeling\label{sec:modeling}}
The search is performed by simultaneously estimating the signal and background yields using a maximum likelihood fit to the data in the 12 event categories. The data are binned in two dimensions, \mhh and \mhbb, with the ranges specified in Section~\ref{sec:selection} and with bin widths of 25 and $2\GeV$, respectively.
The bin widths are smaller than the mass resolutions, but large enough to keep the number of bins computationally tractable.
Each processes is modeled with two-dimensional templates, one for each event category. The templates are created using simulation. Because of the limited size of the simulated samples, we employ methods to smooth the background distributions. Shape uncertainties that account for possible differences between data and simulation are included while executing the fit.  This fitting method was previously presented in Ref.~\cite{Sirunyan:2018iff}.

\subsection{Background categorization}

Background events are separated into four generator-level categories, each with distinct \mhbb shapes. The categories are defined by counting the number of generator-level quarks from the immediate decay of a top quark, $\PW$ boson, or (rarely) $\PZ$ boson within $\DR<0.8$ of the \hbbjet axis. The first, labeled ``\mtbkg,'' is the component in which all three quarks from a single top quark decay fulfill this criterion. The second is labeled ``\mwbkg'' and consists of those events that are not labeled \mtbkg but in which both quarks from a $\PW$ or $\PZ$ boson fall within the jet cone. Both of these backgrounds contain resonant peaks in the \mhbb shape corresponding to either the top quark or $\PW$ boson mass. The ``\losttwbkg'' contains events with partial decays within the \hbbjet, identified as events in which at least one quark is contained within the jet cone, but does not satisfy one of the previous two requirements. The last category, ``\qgbkg,'' designates all other events. The first three categories are primarily composed of \ttbar events, while the last is a composite of \wjets, multijet, and \ttbar events. The background categorization is summarized in Table~\ref{bkg-type}.

\begin{table}[ht]
\centering
\topcaption{\label{bkg-type} The four exclusive background categories with their kinematical properties and defining number of generator-level quarks within $\DR<0.8$ of the \hbbjet axis.}
\begin{tabular}{llll}
	Bkg. category & Dominant SM process(es) & Resonant in \mhbb & Num. of gen.-level quarks\\[1pt]
\hline
\mtn & \ttbar & Yes (near $m_\PQt$) & 3 from $\PQt$\\
\mwn & \ttbar & Yes (near $m_\PW$) & 2 from $\PW$\\
\Losttwn & \ttbar & No & 1 or 2 \\
\qgn & \wjets and multijet & No & 0\\
\hline
\end{tabular}
\end{table}

\newcommand{\pbkg}{\ensuremath{P_{\text{bkg}}}\xspace}
\newcommand{\pbb}{\ensuremath{P_{\bbbar}}\xspace}
\newcommand{\phh}{\ensuremath{P_{\PH\PH}}\xspace}
\newcommand{\psig}{\ensuremath{P_{\text{signal}}}\xspace}
\newcommand{\deltabb}{\ensuremath{\Delta_{\bbbar}}\xspace}
\newcommand{\mubb}{\ensuremath{\mu_{\bbbar}}\xspace}
\newcommand{\sigmabb}{\ensuremath{\sigma_{\bbbar}}\xspace}
\newcommand{\muhh}{\ensuremath{\mu_{\PH\PH}}\xspace}
\newcommand{\sigmahh}{\ensuremath{\sigma_{\PH\PH}}\xspace}

\subsection{Template creation strategy}

A template is produced for each of the 12 event categories, for each of the four backgrounds.  To reduce statistical fluctuations in the templates, each is generated from an initial smooth template created by relaxing requirements or by combining categories. In all cases, the regions with relaxed criteria are chosen such that the shapes for these regions are similar to those for the full event selection.  The final template for each event category and background is produced by fitting the high-statistics template to the simulated samples for that category's event selection. The fit is performed in a similar manner to the fit to data and with a similar parameterization of the template shape.  The templates are compared to simulation after applying the full event selection and any deviations in shape are found to be much smaller than the statistical uncertainty of the data sample. The background templates and associated systematic uncertainties are ultimately validated by fitting to data in dedicated control regions, which  is described in Section~\ref{sec:crtests}.

While this procedure increases the statistical power of the simulation samples, the multijet background simulation sample cannot be produced with a large enough effective integrated luminosity to be directly used in the template creation. Instead, the similarity of \mhbb reconstruction for \wjets and multijet events is exploited. Both these processes have \hbbjets that are composed of at least one quark or gluon that is misidentified as a \hbbjet, resulting in nearly identical monotonically falling \mhbb shapes. Both processes also have similar relative fractions in the \bL, \bM, and \bT categories. The \wjets and multijet samples are used to obtain a combined yield and \mhh distribution for each lepton flavor and \wqqtau category. The \mhbb modeling and the relative \hbbjet subjet {\cPqb} tagging categorization is then taken from the \wjets sample. These two components are combined to form a single background shape when forming the \qgbkg templates.

\subsection{Background process modeling}

The background templates are modeled as conditional probabilities of \mhbb as a function of \mhh so that the templates include the correlation of these two variables. The two-dimensional probability distribution is
\begin{linenomath}
\begin{equation}
\pbkg(\mhbb,\mhh) = \pbb(\mhbb|\mhh,\theta_1) \phh(\mhh|\theta_2),
\end{equation}
\end{linenomath}
where \phh and \pbb are one-dimensional probability distributions and the $\theta_1$ and $\theta_2$ are nuisance parameters used to account for shape uncertainties. A parametric function that models the full \mhh range for background events is difficult to obtain from first principles. Instead, a non-parametric approach is taken. The \phh are produced from the one-dimensional \mhh histograms with kernel density estimation (KDE)~\cite{Rosenblatt,Silverman,Cranmer:2000du}. The smoothing of the \phh distributions is controlled by parameters within the KDE framework called bandwidths. Gaussian kernels with adaptive bandwidths are used because the event density for this distribution varies strongly with \mhh and a single, global bandwidth is not suitable for the full distribution. These adaptive bandwidths depend on a first iteration estimate of \phh, which itself is produced with KDE. However, for this first iteration a global bandwidth $h$ is used that scales as
\begin{linenomath}
\begin{equation}
h \propto \left(\frac{(\sum_{i=1}^n w_{i})^2}{\sum_{i=1}^n w_{i}^2}\right)^{-1/5}.
\end{equation}
\end{linenomath}
The sums are over all events in the simulation sample and the $w_i$ are the individual event weights. This formulation is chosen to minimize the mean integrated squared error of the estimate. For the adaptive estimates, the bandwidths $h_i$ associated with each event are
\begin{linenomath}
\begin{equation}
h_i=h\ \left(\frac{g}{\widetilde{f}(x_i)}\right)^{1/2},
\end{equation}
\end{linenomath}
where the $\widetilde{f}(x_i)$ are the estimated event densities at the location $x_i$ of the event and $g$ is a normalization factor such that the global bandwidth scale is controlled by $h$. As discussed in Ref.~\cite{Scott}, adaptive KDE can result in overestimation of the distribution tails in the case of large bandwidths being applied. This is ameliorated by imposing a maximum bandwidth value, which is usually chosen to be 1--5 times larger than the median bandwidth. The \mhh tail is further smoothed by fitting with an exponential function for $\mhh\gtrsim 2\TeV$.

The \pbb distributions are obtained for the \mtn and \mwn backgrounds by fitting \mhbb histograms with a double Crystal Ball function~\cite{Oreglia:1980cs,Gaiser:1982yw}.
This function has a Gaussian core, which is used to model the bulk of the \mhbb distribution, and power-law tails, which describe the effects of more severe jet misreconstruction.
The fits are performed for events binned in \mhh to capture the evolution of the \mhbb shape with \mhh. The double Crystal Ball function parameters are then interpolated between \mhh bins. The \pbb distributions for the \losttwn and \qgn backgrounds are estimated from the two-dimensional histograms with two-dimensional KDE. Independent adaptive bandwidths and bandwidth upper limits are used for each dimension when forming the \pbb. Similar to the derivation of the \phh, the \mhh tails are smoothed with exponential function fits. Simulation yields are used as the initial values of the background yields in the fit to data.

\subsection{Signal process modeling}

The signal templates are also modeled as conditional probabilities
\begin{linenomath}
\begin{equation}
\psig(\mhbb,\mhh|\mx) = \phh(\mhh|\mhbb,\mx,\theta_1) \pbb(\mhbb|\mx,\theta_2).
\end{equation}
\end{linenomath}
The \psig distributions are first obtained for discrete \mx values by fitting histograms of the signal mass distributions. Models continuous in \mx are then produced by interpolating the fit parameters. The \pbb distributions are created by fitting \mhbb histograms with a double Crystal Ball function, and the resonance resolution is ${\approx}10\%$. The shape for the \bL categories also includes an exponential function to model the small fraction of signal events with no resonant peak in the distribution.

The \phh distributions are also modeled with a double Crystal Ball function, but with a linear dependence on \mhbb, parameterized by $\deltabb = (\mhbb - \mubb)/\sigmabb$. The \mubb and \sigmabb are the mean and standard deviation parameters from the fit to \mhbb, respectively. 
The variable \muhh, the mean of the Crystal Ball function, is then
\begin{linenomath}
\begin{equation}
    \muhh = \mu_0 (1 + \mu_1  \deltabb),
\end{equation}
\end{linenomath}
where $\mu_0$ and $\mu_1$ are fit parameters. This parameterization models the characteristic that a mismeasurement of the \hbbjet results in a mismeasurement of \mhh. The standard deviation of \mhh, denoted as \sigmahh, also depends on \mhbb,
\begin{linenomath}
\begin{equation}
  \sigmahh =
    \begin{cases}
      \sigma_0 (1+\sigma_1 \abs{\deltabb}), & \deltabb < 0, \\
      \sigma_0, & \deltabb > 0, \\
    \end{cases}
\end{equation}
\end{linenomath}
where $\sigma_0$ and $\sigma_1$ are fit parameters. An undermeasurement of \mhbb can be caused by the SD algorithm removing energy from the Higgs boson decay. In such a scenario, the correlation between the two variables worsens and the \mhh resolution becomes wider. For $\abs{\deltabb}>2.5$, only the values at the boundary are used since the correlation does not hold for severe mismeasurements. The \mhh resolution is ${\approx}6\%$ for $\mx=1\TeV$, decreasing to $4\%$ for $\mx=3\TeV$.

The product of the acceptance and efficiency for $\PX\to\PH\PH$ events to be included in the individual event categories is taken from simulation. As for the shape parameters, the efficiency is interpolated in \mx. Uncertainties in the relative acceptances and in the integrated luminosity of the sample are included in the maximum likelihood fit that is used to obtain confidence intervals on the $\PX\to\PH\PH$ process. The modeling is tested by fitting the templates to pseudo-experiments with injected signal and no significant bias in the fitted signal yield is found.

\subsection{Validation of background models with control region data\label{sec:crtests}}

\begin{figure}[ht!]
\centering
\includegraphics[width=0.45\textwidth]{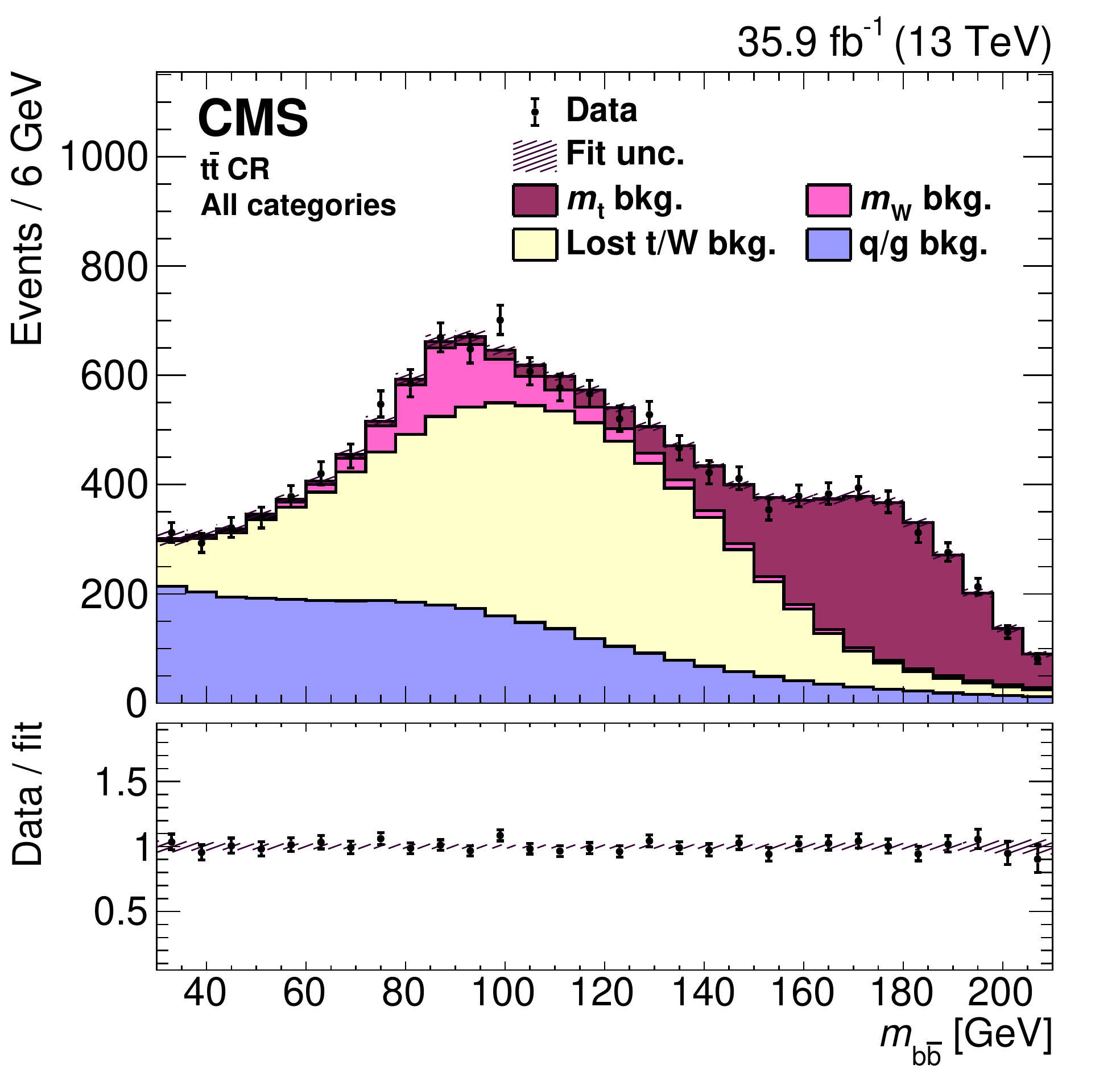}
\includegraphics[width=0.45\textwidth]{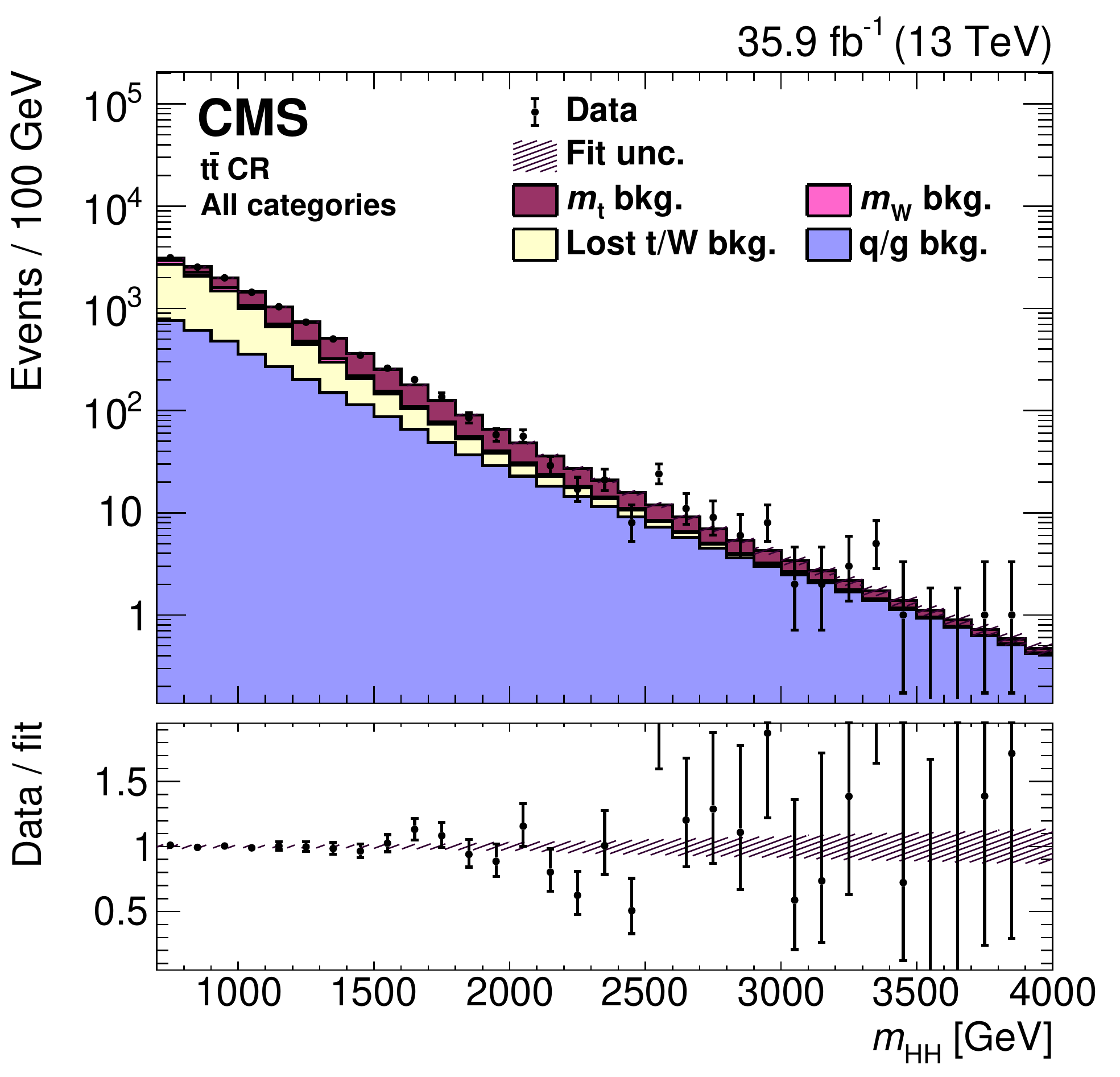}
\includegraphics[width=0.45\textwidth]{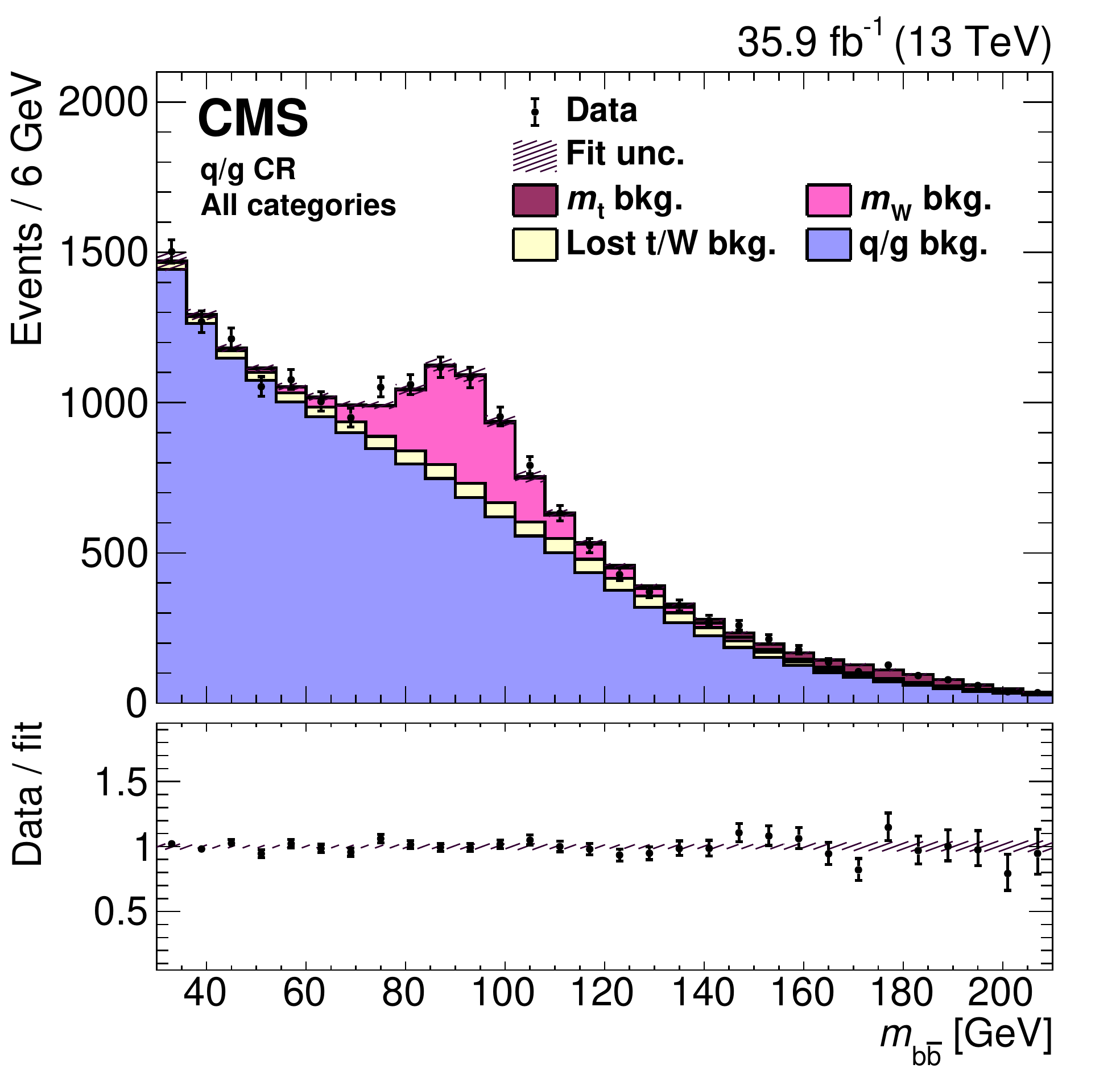}
\includegraphics[width=0.45\textwidth]{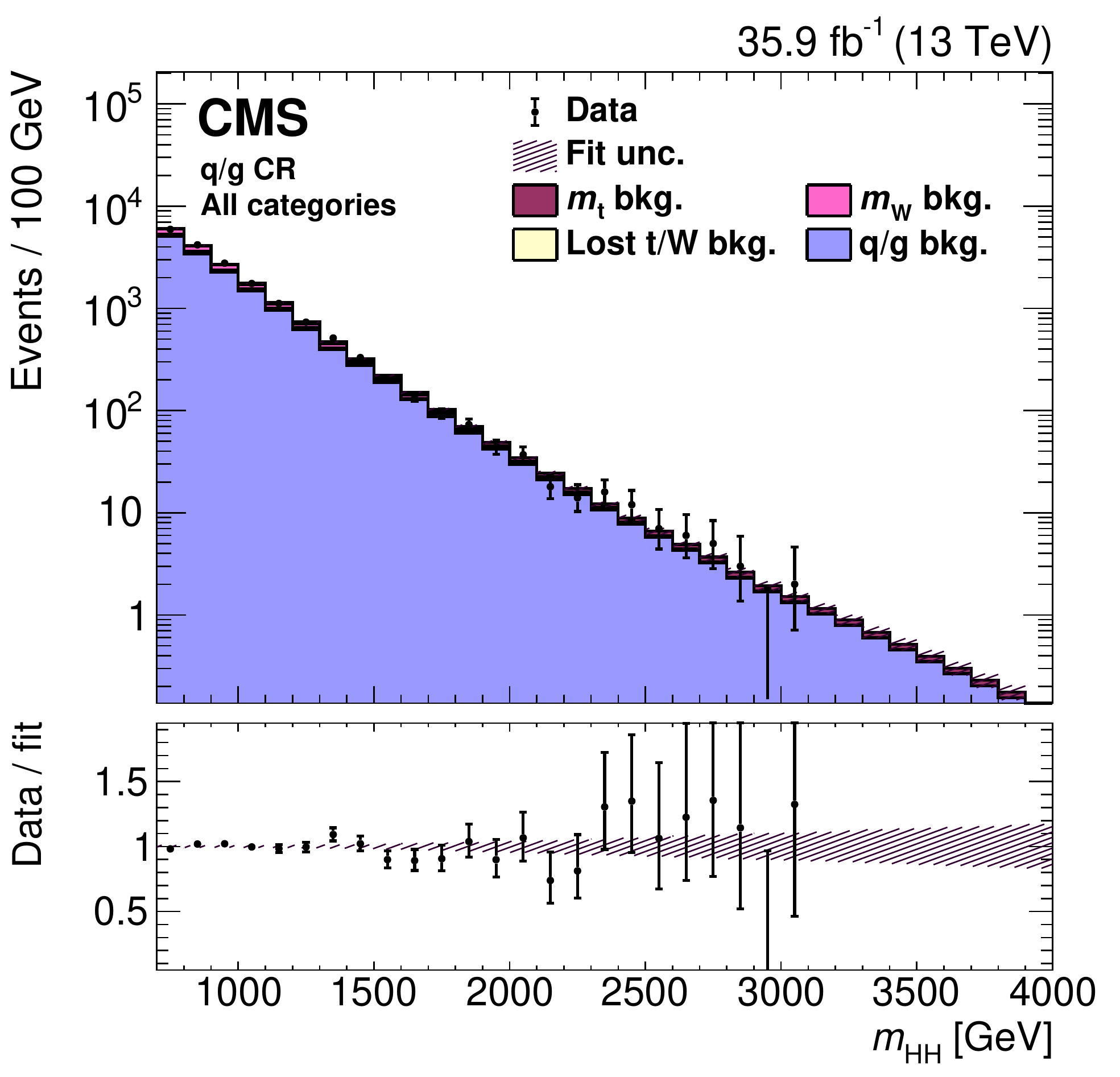}

\caption{The fit result compared to data in the $\ttbar$~CR (upper plots) and \qgn~CR (lower plots), projected in \mhbb (left) and \mhh (right). Events from all categories are combined. The fit result is the filled histogram, with the different colors indicating different background categories. The background shape uncertainty is shown as the hatched band. The lower panels show the ratio of the data to the fit result.}
\label{fig:modeling_cr}
\end{figure}

The background models are validated by analyzing the $\ttbar$~CR and \qgn~CR data samples. For both control regions, background templates are constructed in the same way as for the standard event selection, except that they are made to model the control region selection. The background templates are then fit to the control region data with the same systematic uncertainties that are used in the standard maximum likelihood fit. The result of the simultaneous fit is shown in Fig.~\ref{fig:modeling_cr} for both control regions.
To improve visualization, the displayed binning shown in this and subsequent figures is coarser than that used in the maximum likelihood fit.
The projections in both mass dimensions are shown for the combination of all event categories. The fit result models the data well, indicating that the shape uncertainties can account sufficiently for potential differences between data and simulation.

\section{Systematic uncertainties\label{sec:syst}}
Systematic uncertainties are included in the maximum likelihood fit as nuisance parameters. Nuisance parameters for shape uncertainties are modeled as Gaussian functions, whereas log-normal functions are used for normalization uncertainties. The \mhbb scale and resolution uncertainties for the signal, the \mtbkg, and the \mwbkg are evaluated as uncertainties in the mean and standard deviation of the double Crystal Ball function parameters, respectively. The signal \mhh scale and resolution uncertainties are handled in the same manner. The other background shape uncertainties are implemented as alternative background templates. Each alternative template is produced by shifting the nominal background template, bin-by-bin, by a factor that depends on either \mhh or \mhbb. The magnitudes of these factors are subsequently constrained as nuisance parameters. 

The parameterization of the background uncertainties is motivated by the expectation of possible differences between simulation and data for such aspects
as background composition or jet energy scale. Studies of the $\ttbar$~CR and the \qgn~CR are used to verify that the chosen uncertainties do cover these differences. More complex background models, such as those with more nuisance parameters or higher order shape distortions, are also tested in these control regions. The more complex background models do not lead to better agreement between data and the fit result. The fit result does not depend strongly on the initial uncertainty sizes because they function only as loose constraints for the fit. This is verified by inflating all initial background uncertainty sizes by a factor of two and observing that the final result does not change. Therefore, the initial background uncertainty sizes are sufficiently large to easily account for the differences between simulation and data in the control regions. 

Shape distortions derived from differences between simulation generator programs, parton showering and simulation programs, and matrix element calculation order were also studied. The uncertainties used in obtaining this result are comparable to or larger than those derived from these differences. Each uncertainty is listed in Table~\ref{systematics} with its initial size. A single uncertainty type can be applied to multiple event categories with independent nuisance parameters per category. The background model includes 98 nuisance parameters, while the signal model includes 13 and shares an additional two with the background model. The description of each uncertainty, including correlations between event categories, is described in Sections~\ref{sec:syst_bkg}--\ref{sec:syst_signal}.

\begin{table}[ht]
\topcaption{\label{systematics} The systematic uncertainties included in the maximum likelihood fit and how they are applied to each process model. The ``type'' indicates if the uncertainty affects process yield $Y$ or the shape of the \mhbb or \mhh distributions. Some uncertainties are applied to multiple event categories with independent nuisance parameters. The number of such parameters, $N_\text{p}$, the initial uncertainty size, $e_{\text{I}}$, and the ratios of the constrained size to the initial size, $e_{\text{C}}/e_{\text{I}}$, are listed. The ratios are obtained by fitting a model containing only background processes to the data. Uncertainty sizes that vary by event category are listed with category labels. The labels $Y$, $S$, and $R$ denote how a single uncertainty affects yield, scale, and resolution, respectively.} 
\begin{tabular}{llllll}
Uncertainty label & Type & Processes &$N_\text{p}$ & $e_{\text{I}}$ & $e_{\text{C}}/e_{\text{I}}$   \\
\hline
\qgn normalization              & $Y$       & \qgn                & 12 & 50\%                              & 27--48\%  \\
\Nonqg normalization            & $Y$       & \mtn, \mwn, \losttwn& 12 & 25\%                              & 31--85\%  \\
\Nonqg categorization           & $Y$       & \mtn, \mwn, \losttwn& 6  & 25\%                              & 12--99\%  \\
\Nonqg cat. \pt dep.            & \mhh      & \mtn, \mwn, \losttwn& 6  & $\pm0.13(\mhh/\TeV)$              & 91--99\%  \\[\cmsTabSkip]
SD scale                        & \mhbb     & \mtn, \mwn, signal  & 1  & 1\%                               & 52\%      \\
SD resolution                   & \mhbb     & \mtn, \mwn, signal  & 1  & 20\%                              & 31\%      \\
\Losttwn \mhbb scale            & \mhbb     & \Losttwn            & 3  & $\pm0.0015(\mhbb/\GeV)$           & 91--99\%  \\
\Losttwn low \mhbb              & \mhbb     & \Losttwn            & 3  & $\pm18(\GeV/\mhbb)$               & $>$87\%   \\
\qgn \mhbb scale                & \mhbb     & \qgn                & 3  & $\pm0.0025(\mhbb/\GeV)$           & 90--96\%  \\
\qgn low \mhbb                  & \mhbb     & \qgn                & 3  & $\pm30(\GeV/\mhbb)$               & 40--60\%  \\[\cmsTabSkip]
\Nonqg \mhh scale               & \mhh      & \mtn, \mwn, \losttwn& 12 & $\pm0.13(\mhh/\TeV)$              & 94--99\%  \\
\Nonqg \mhh resolution          & \mhh      & \mtn, \mwn, \losttwn& 12 & $\pm0.28(\TeV/\mhh)$              & 95--99\%  \\
\qgn \mhh scale                 & \mhh      & \qgn                & 12 & $\pm0.5(\mhh/\TeV)$               & 77--96\%  \\
$\qgn$ \mhh resolution          & \mhh      & \qgn                & 12 & $\pm1.4(\TeV/\mhh)$               & 58--87\%  \\[\cmsTabSkip]
Luminosity                      & $Y$       & Signal              & 1  & 2.5\%                             & \NA       \\
PDF and scales                  & $Y$       & Signal              & 1  & 2\%                               & \NA       \\
Trigger                         & $Y$       & Signal              & 2  & 2\%                               & \NA       \\
Lepton selection                & $Y$       & Signal              & 2  & \Pe:$5.7\%$ $\mu$:$5.3\%$         & \NA       \\
Jet energy scale                & $Y,\mhh$  & Signal              & 1  & $Y$:$0.5\%$ $S$:$1\%$ $R$:$2\%$   & \NA       \\
Jet energy res.                 & $Y,\mhh$  & Signal              & 1  & $Y$:$1\%$ $S$:$0.5\%$ $R$:$5\%$   & \NA       \\
Unclustered energy              & $Y,\mhh$  & Signal              & 1  & $Y$:$1\%$ $S$:$0.5\%$ $R$:$0.5\%$ & \NA       \\
\hbbjet~{\cPqb} tagging         & $Y$       & Signal              & 1  & ${<}10\%$                         & \NA       \\
AK4 jet {\cPqb} tagging veto    & $Y$       & Signal              & 1  & $1\%$                             & \NA       \\
\wqqtau                         & $Y$       & Signal              & 1  & HP:$14\%$ LP:$33\%$               & \NA       \\
\wqqtau extrapolation           & $Y$       & Signal              & 1  & ${<}7\%$                          & \NA       \\
\hline
\end{tabular}
\end{table}

\subsection{Background normalization uncertainties\label{sec:syst_bkg}}

Since the main source of the \mtn, \mwn, and \losttwn backgrounds is \ttbar production, some uncertainties are applied by treating the three categories as a single component, referred to collectively as the ``\nonqg background.''

The fraction of each of the three categories within the combination is determined from the overall {\cPqb} tagging efficiency and the \hbbjet \pt distributions. Additional uncertainties are then assigned to the modeling of their relative composition.

For each event category, the \qgbkg and the \nonqg background each have a large initial normalization uncertainty that is uncorrelated among categories. The relative composition of the three \ttbar backgrounds is controlled in two ways. First, the \mwn and \losttwn backgrounds have independent normalization uncertainties per {\cPqb} tagging category. In both cases, the \mtbkg normalization is varied in an anticorrelated manner such that the \nonqg background normalization does not change. Second, the composition is allowed to vary linearly with \mhh to account for \hbbjet reconstruction effects that depend on \hbbjet \pt. This is implemented with a \mhh shape uncertainty that only shifts the \mtn background spectrum. There is one such independent nuisance parameter per {\cPqb} tagging category.  Three other nuisance parameters shift the \mwn and \losttwn backgrounds together.

\subsection{Background shape uncertainties}

The jet mass scale and resolution after applying the SD algorithm are measured for \PW~boson decays merged into single jets in data with \ttbar events, using the known $\PW$ boson mass.
The mass scale and resolution in the simulation are found to agree with the data within uncertainties.
These measurements determine the uncertainties in the \mhbb scale and resolution of the \mtn and \mwn backgrounds. For the \losttwn and \qgn backgrounds, nuisance parameters are used to account for mismodeling of the simulated energy scale or the low-mass region by morphing the template shapes using a factor that is either proportional to, or inversely proportional to \mhbb, respectively. The \mhbb shapes do not vary strongly with lepton flavor or \wqqtau, so a single pair of uncorrelated nuisance parameters is applied per background and {\cPqb} tagging category.

Mismodeling of the background \pt spectrum could manifest as an incorrect \mhh scale. This is accounted for by morphing the background templates by multiplicative factors proportional to \mhh. Possible mismodeling of the \mhh resolution is considered in a similar manner, but with multiplicative factors proportional to $\mhh^{-1}$. A pair of scale and resolution uncertainties is assigned to the \nonqg background spectrum for each event category. An independent set of \mhh uncertainties for the \qgbkg is also included.

\subsection{Signal uncertainties\label{sec:syst_signal}}

A 2.5\% uncertainty in the integrated luminosity~\cite{CMS-PAS-LUM-17-001} is included as a signal normalization uncertainty. Signal acceptance uncertainties from the choices of PDF, factorization scale, and renormalization scale are also applied. The scale uncertainties are obtained following the prescription found in Refs.~\cite{Cacciari:2003fi,Catani:2003zt}, and the PDF uncertainty is evaluated using the NNPDF~3.0 PDF set~\cite{Ball:2011mu}. Both the simulated trigger selection efficiency and the lepton selection efficiencies are corrected to match the data efficiencies. The uncertainties in these measurements are included as independent uncertainties in the electron and muon channel signal yields. Uncertainties in the jet energy scale, resolution, and unclustered energy resolution affect signal acceptance, \mhh scale, and \mhh resolution. The same \mhbb scale and resolution uncertainties that are applied to the \mtn and \mwn backgrounds are applied to the signal. In this case, the background and signal uncertainties are $100\%$ correlated.

\begin{figure}[htp!]
\centering
\includegraphics[width=0.45\textwidth]{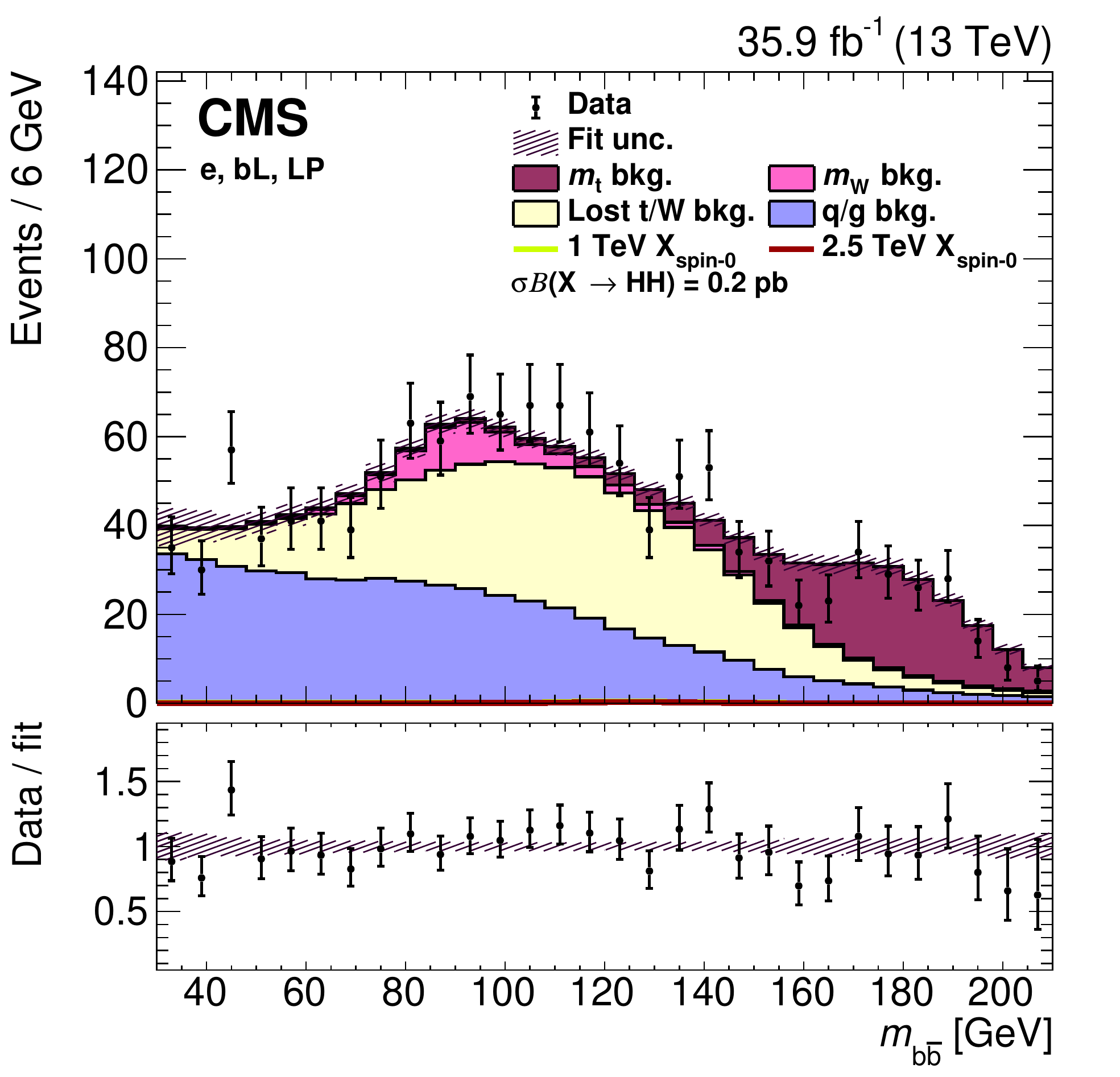}
\includegraphics[width=0.45\textwidth]{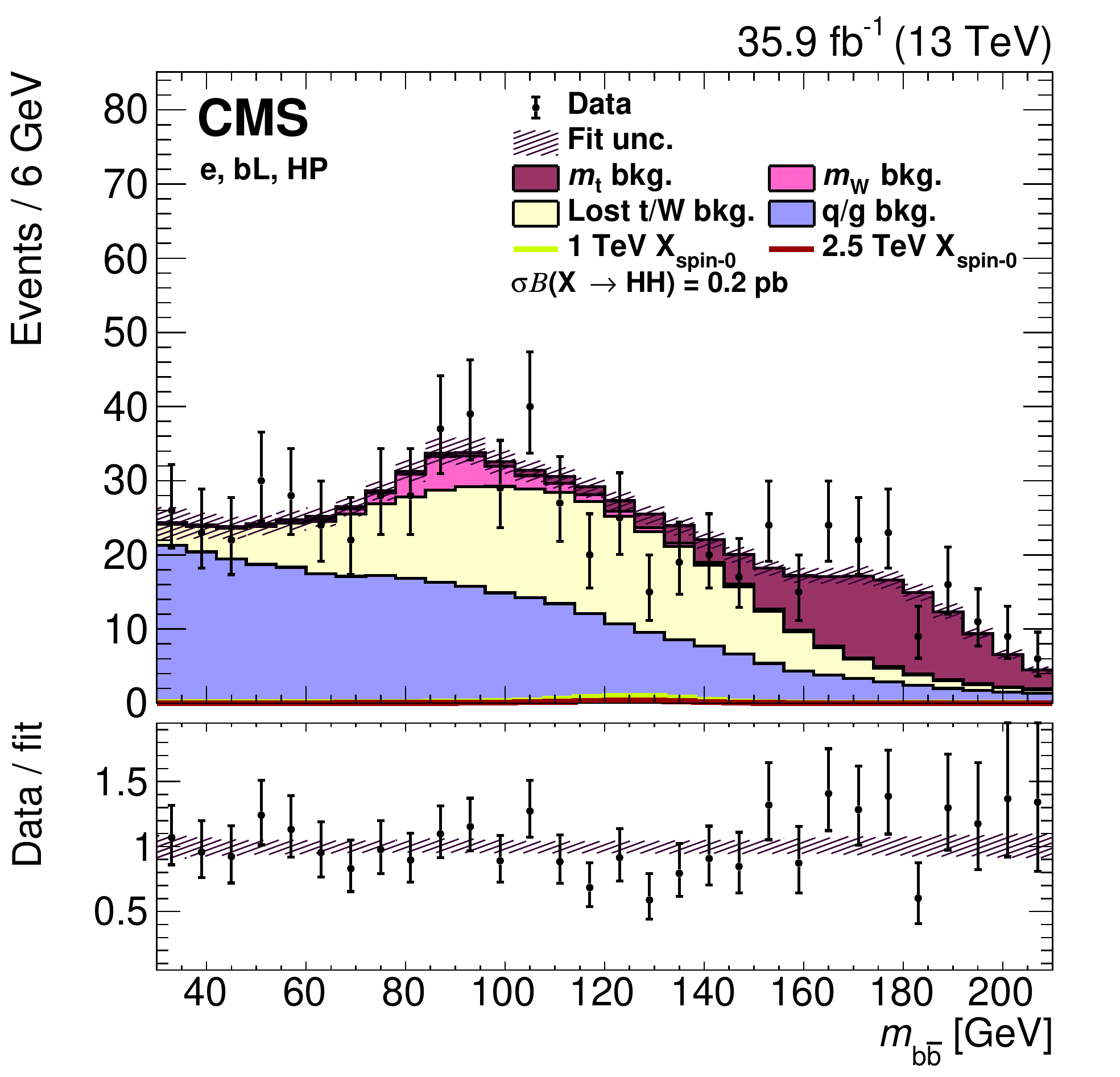}
\includegraphics[width=0.45\textwidth]{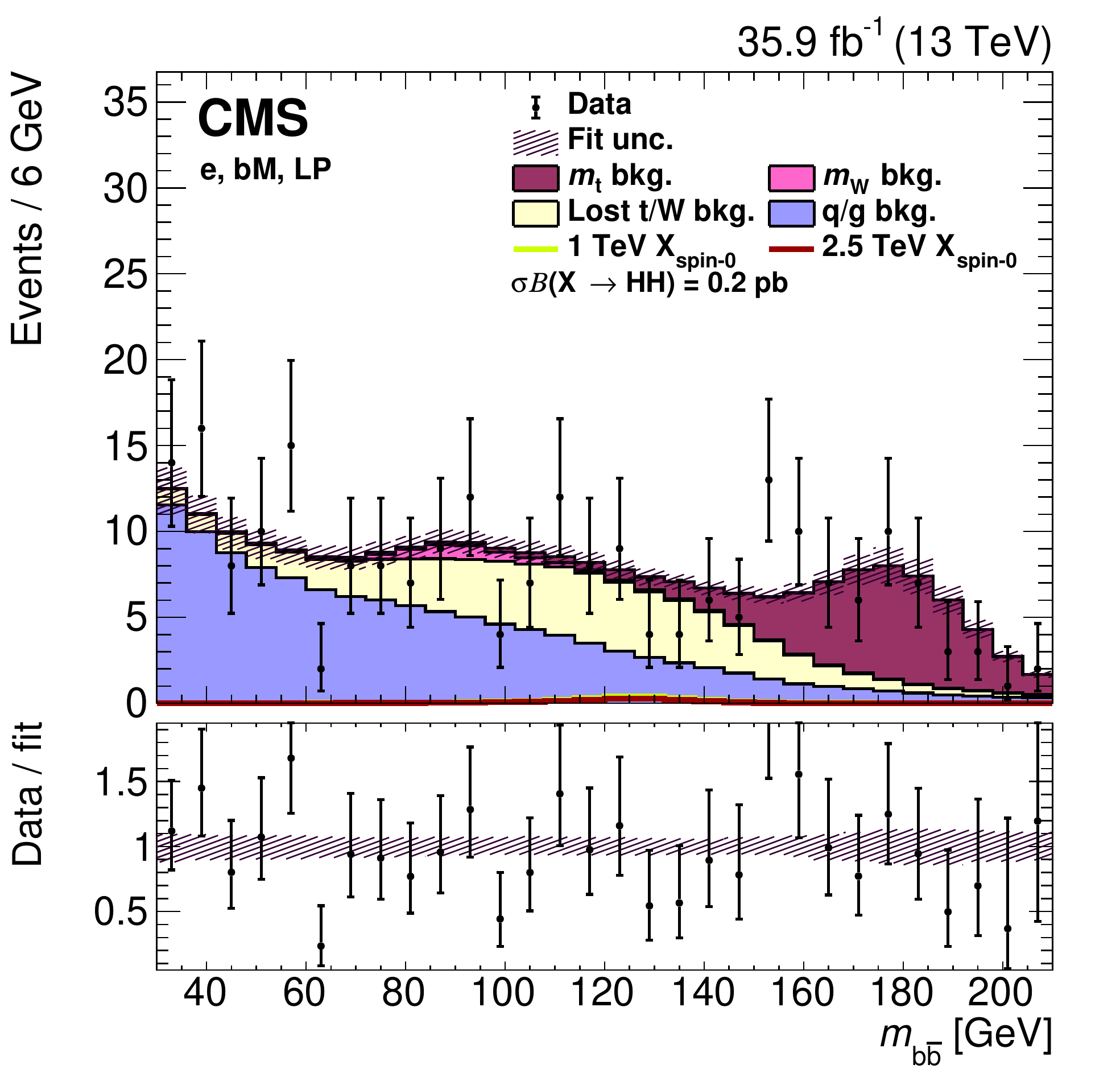}
\includegraphics[width=0.45\textwidth]{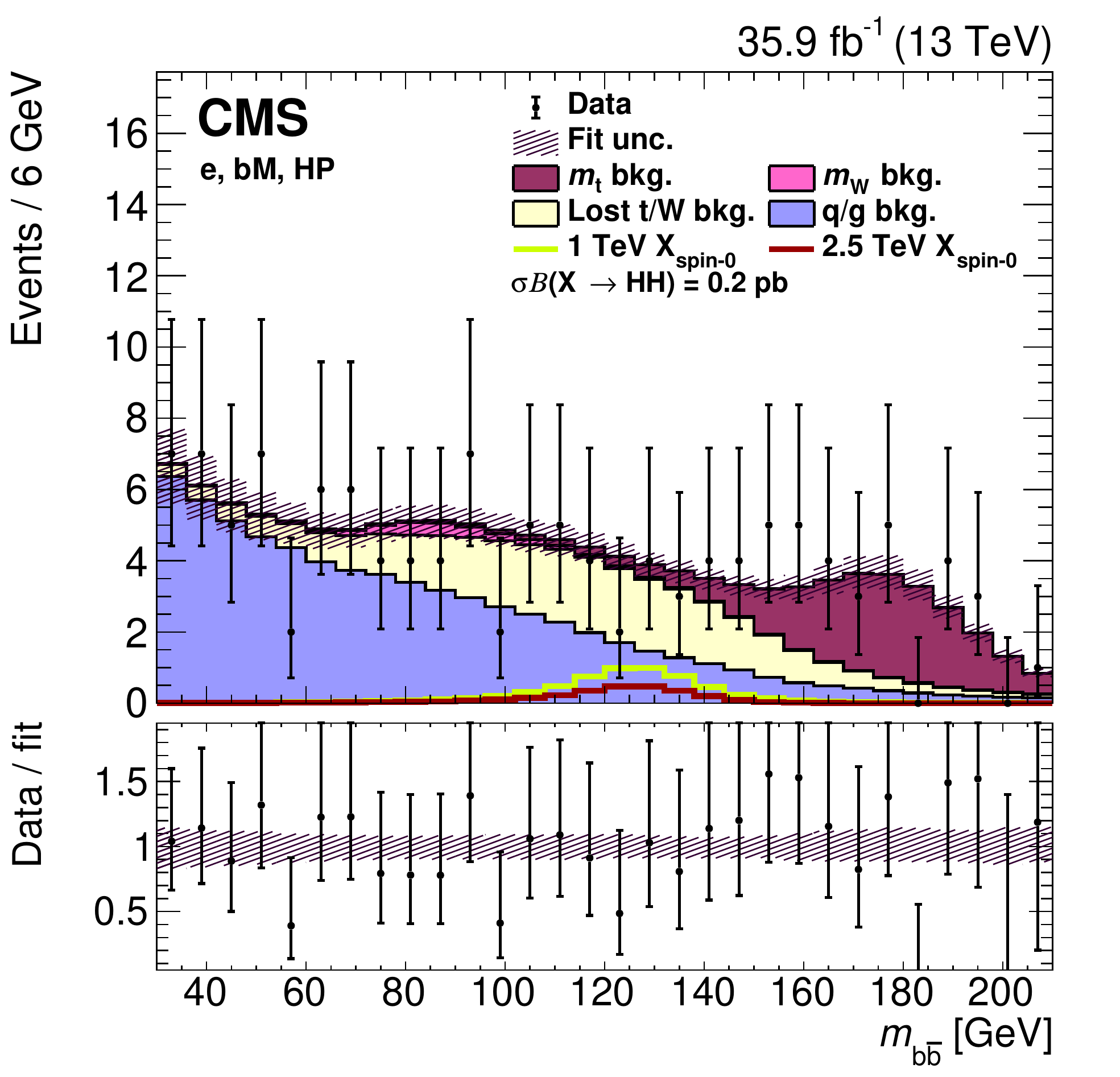}
\includegraphics[width=0.45\textwidth]{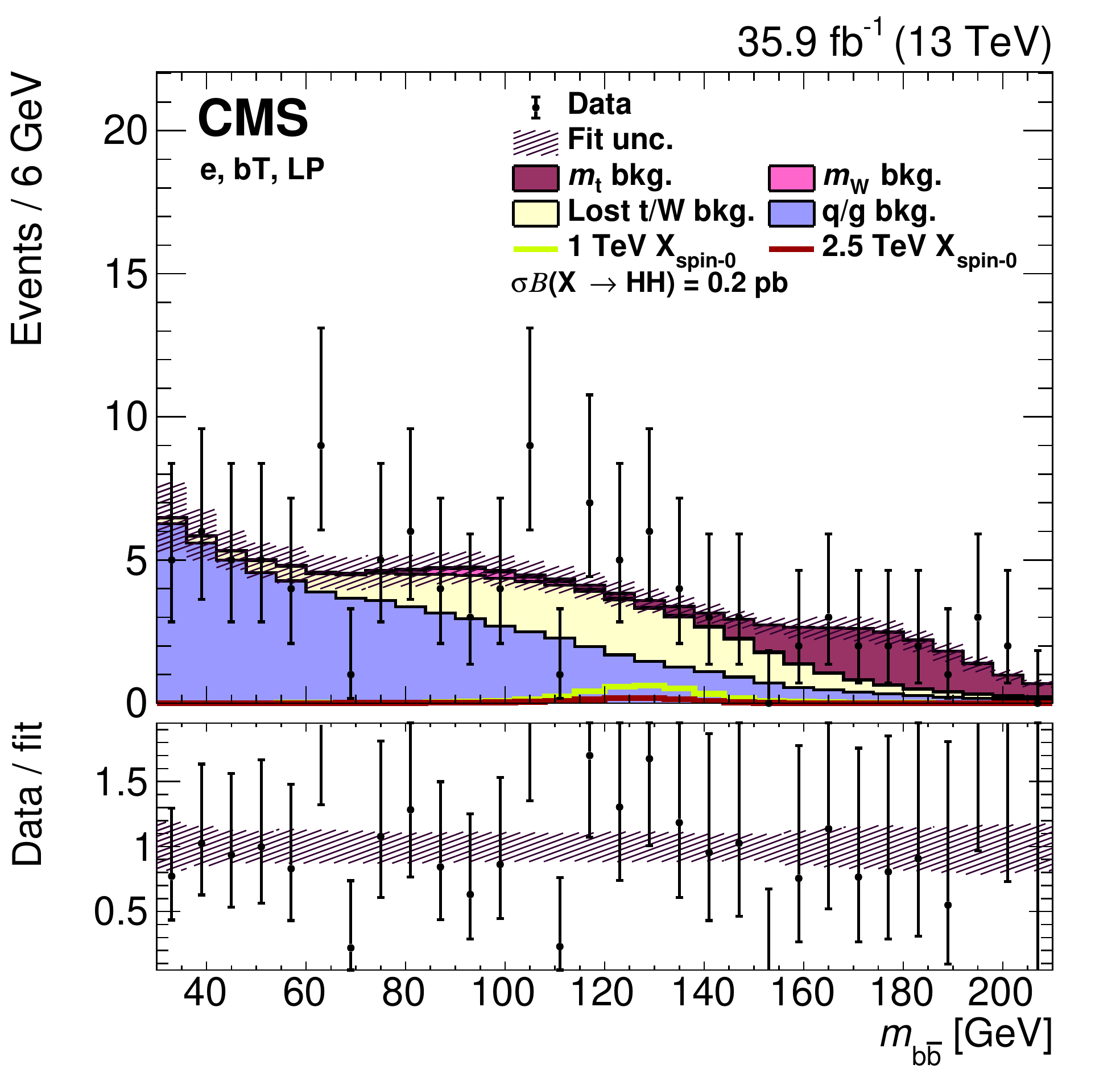}
\includegraphics[width=0.45\textwidth]{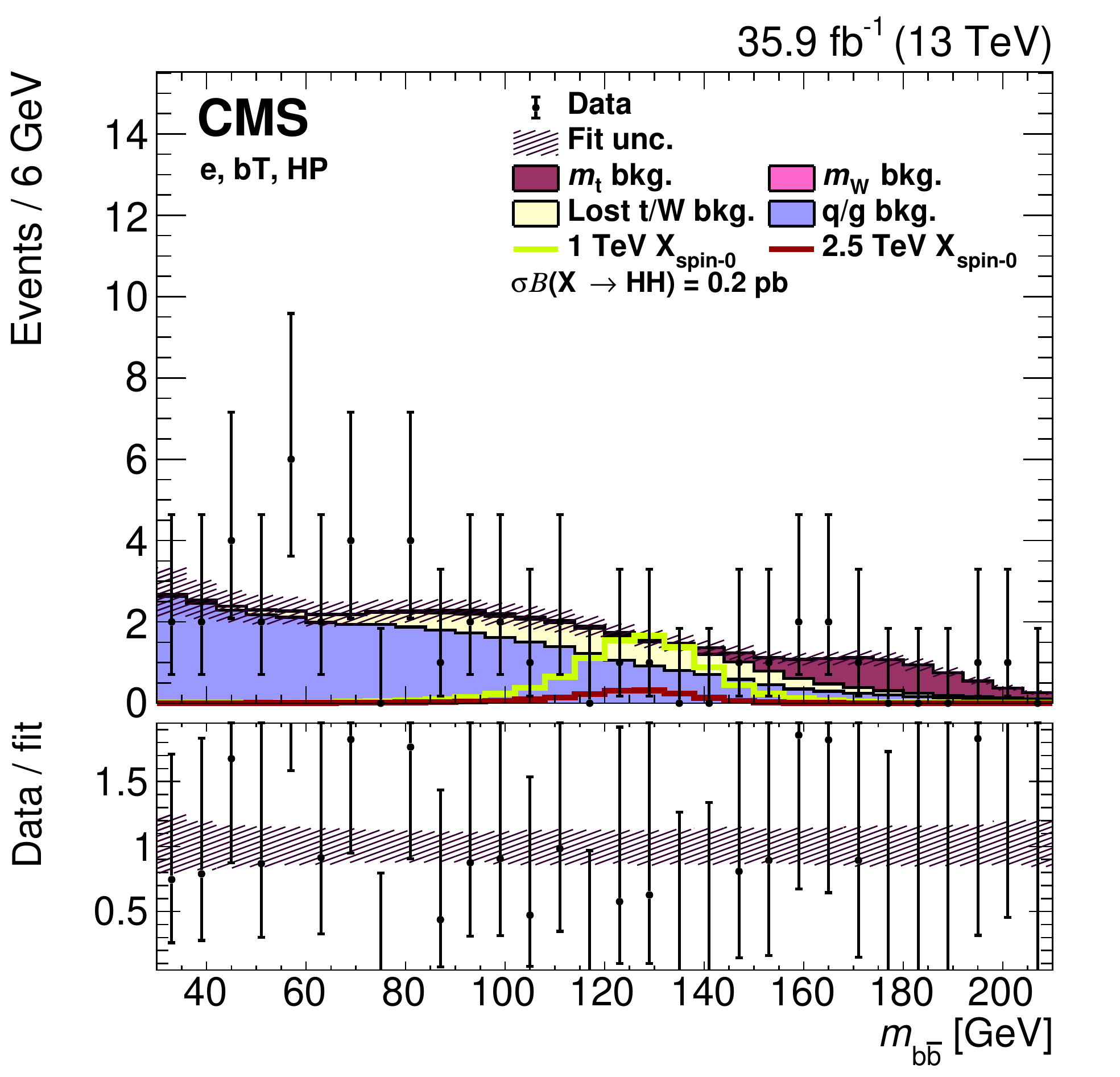}

\caption{The fit result compared to data projected in \mhbb for the electron event categories. The fit result is the filled histogram, with the different colors indicating different background categories. The background shape uncertainty is shown as the hatched band. Example spin-0 signal distributions for $\mx$ of 1 and 2.5\TeV are shown as solid lines, with the product of the cross section and branching fraction to two Higgs bosons set to 0.2\unit{pb}. The lower panels show the ratio of the data to the fit result.}
\label{fig:results_e_mhbb}
\end{figure}
\begin{figure}[htp!]
\centering
\includegraphics[width=0.45\textwidth]{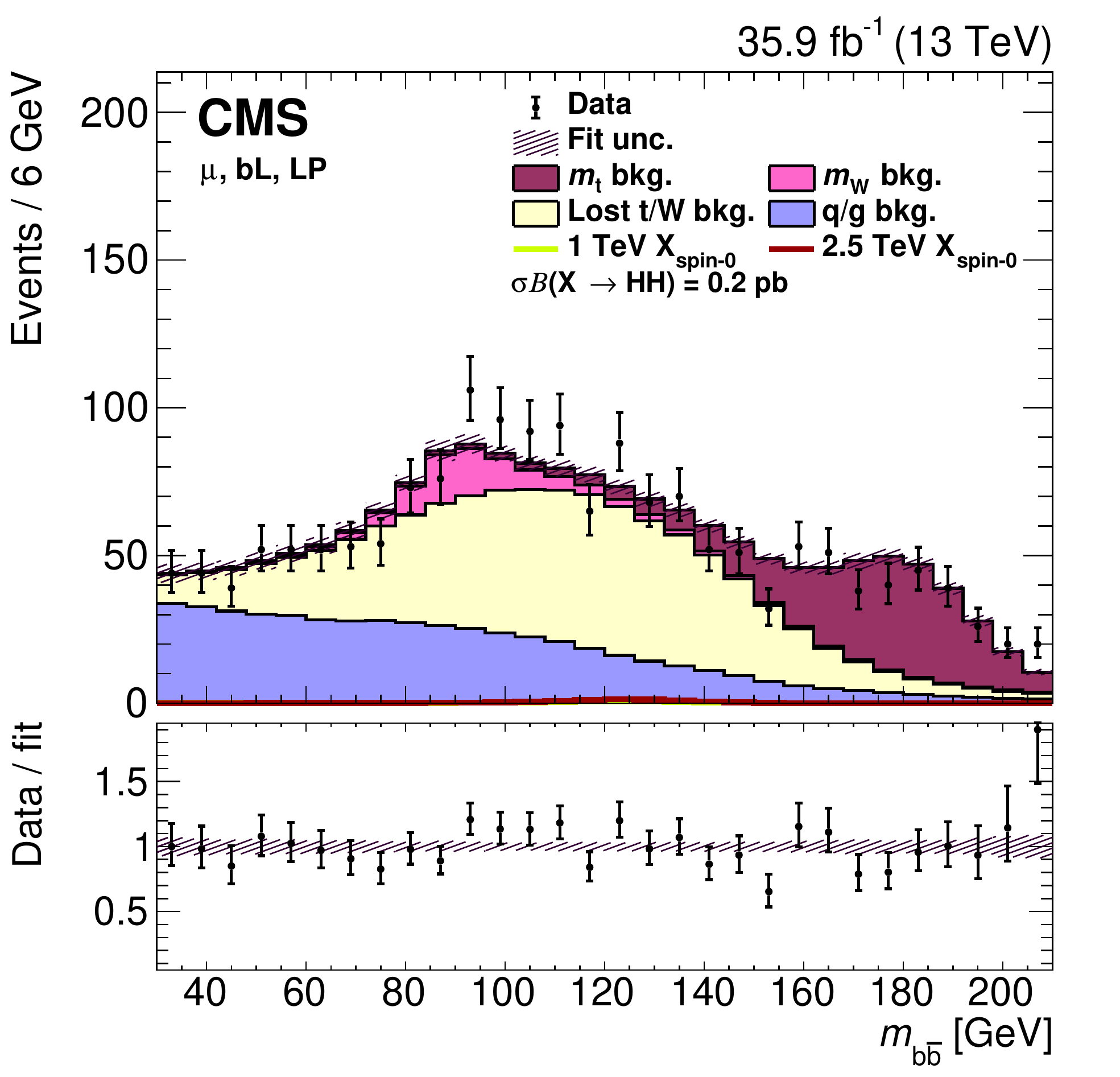}
\includegraphics[width=0.45\textwidth]{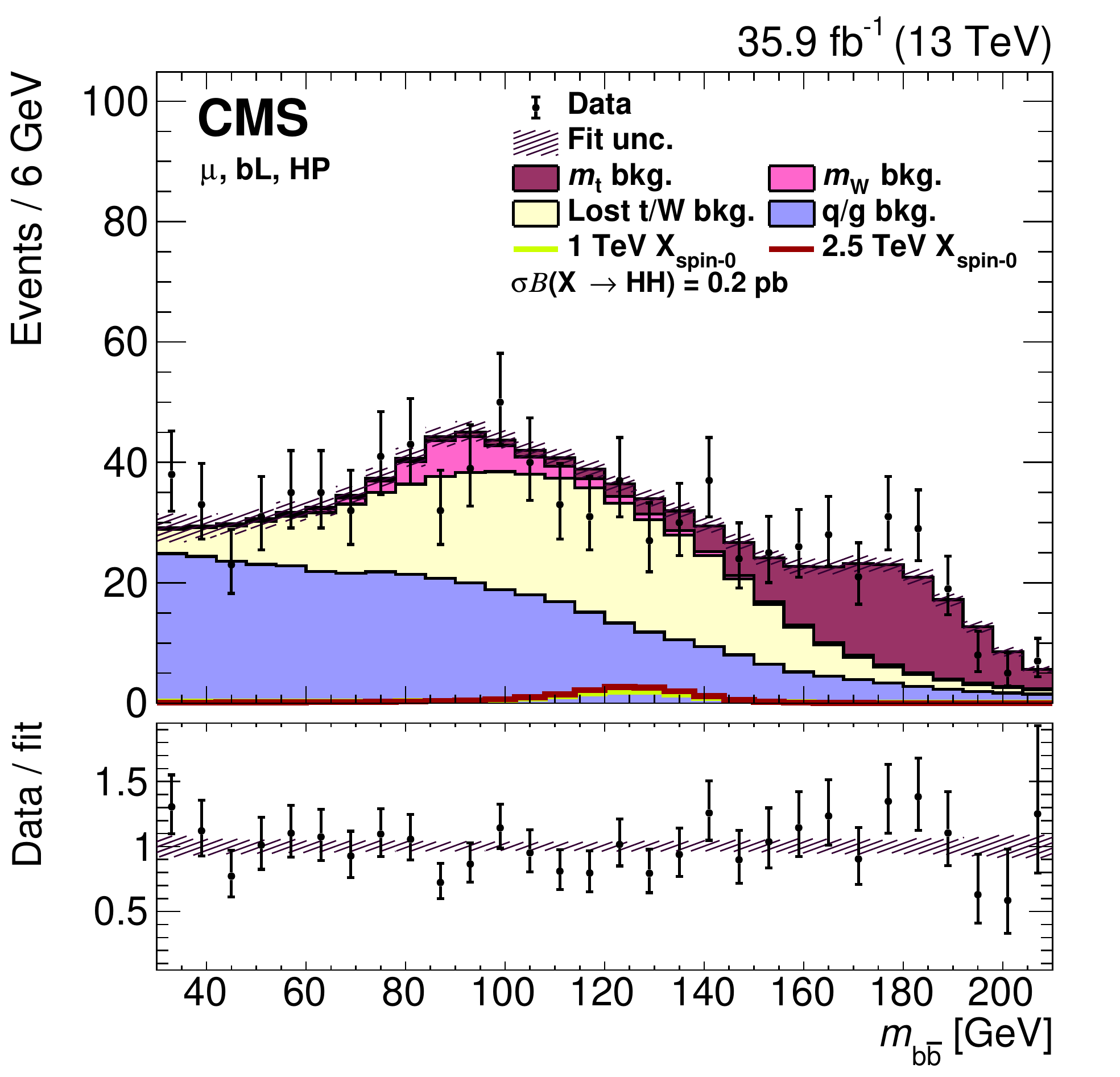}
\includegraphics[width=0.45\textwidth]{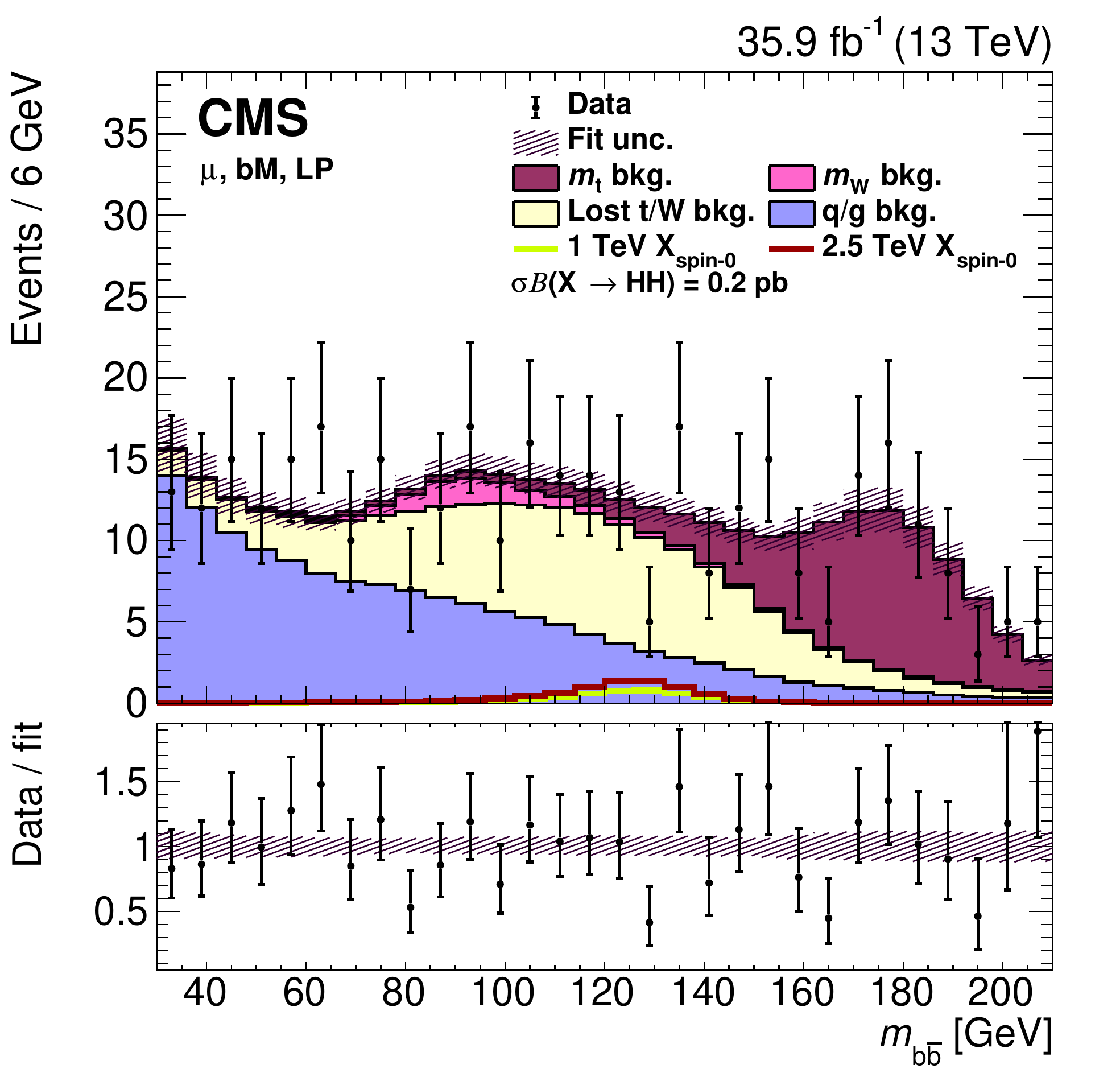}
\includegraphics[width=0.45\textwidth]{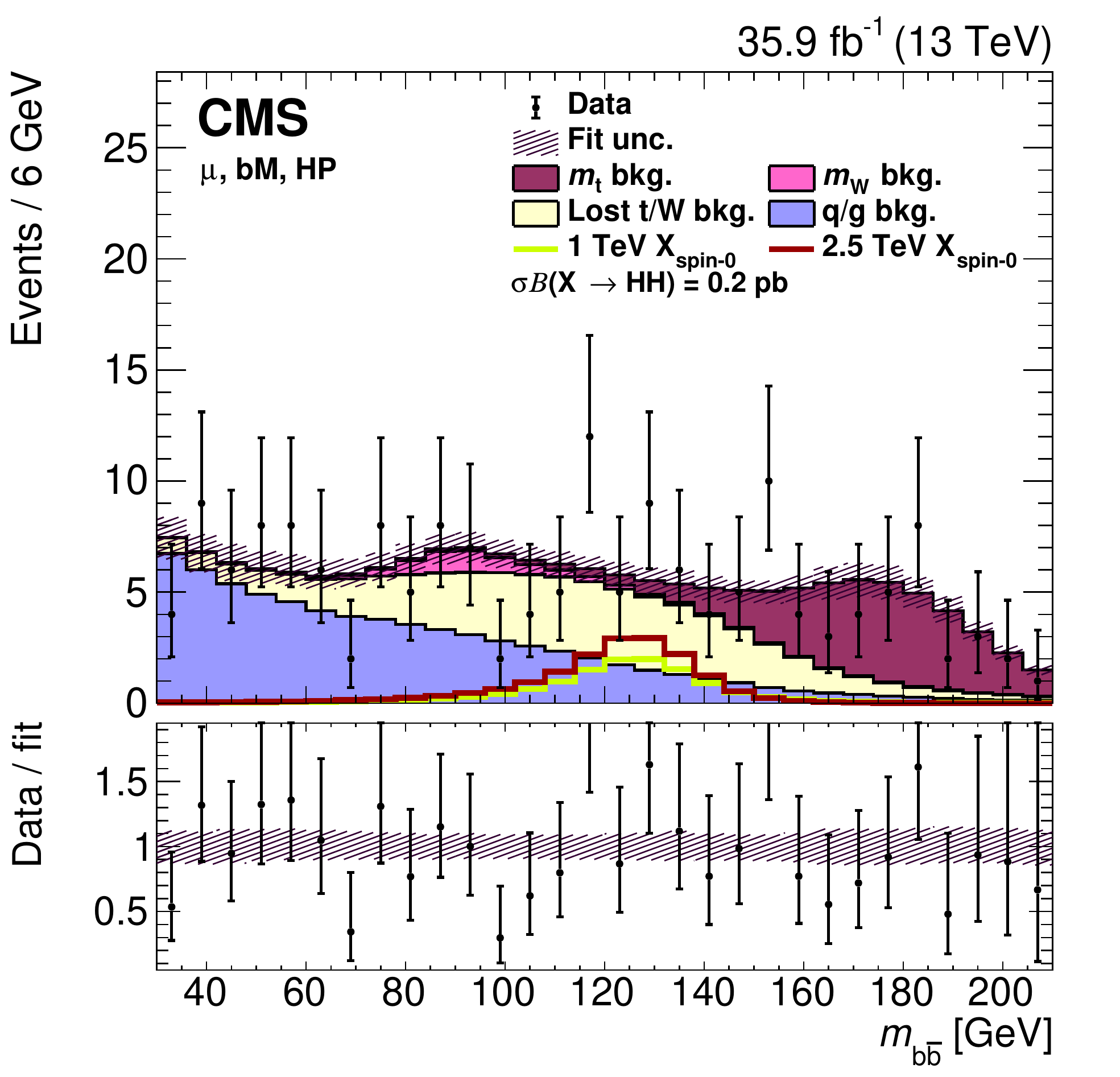}
\includegraphics[width=0.45\textwidth]{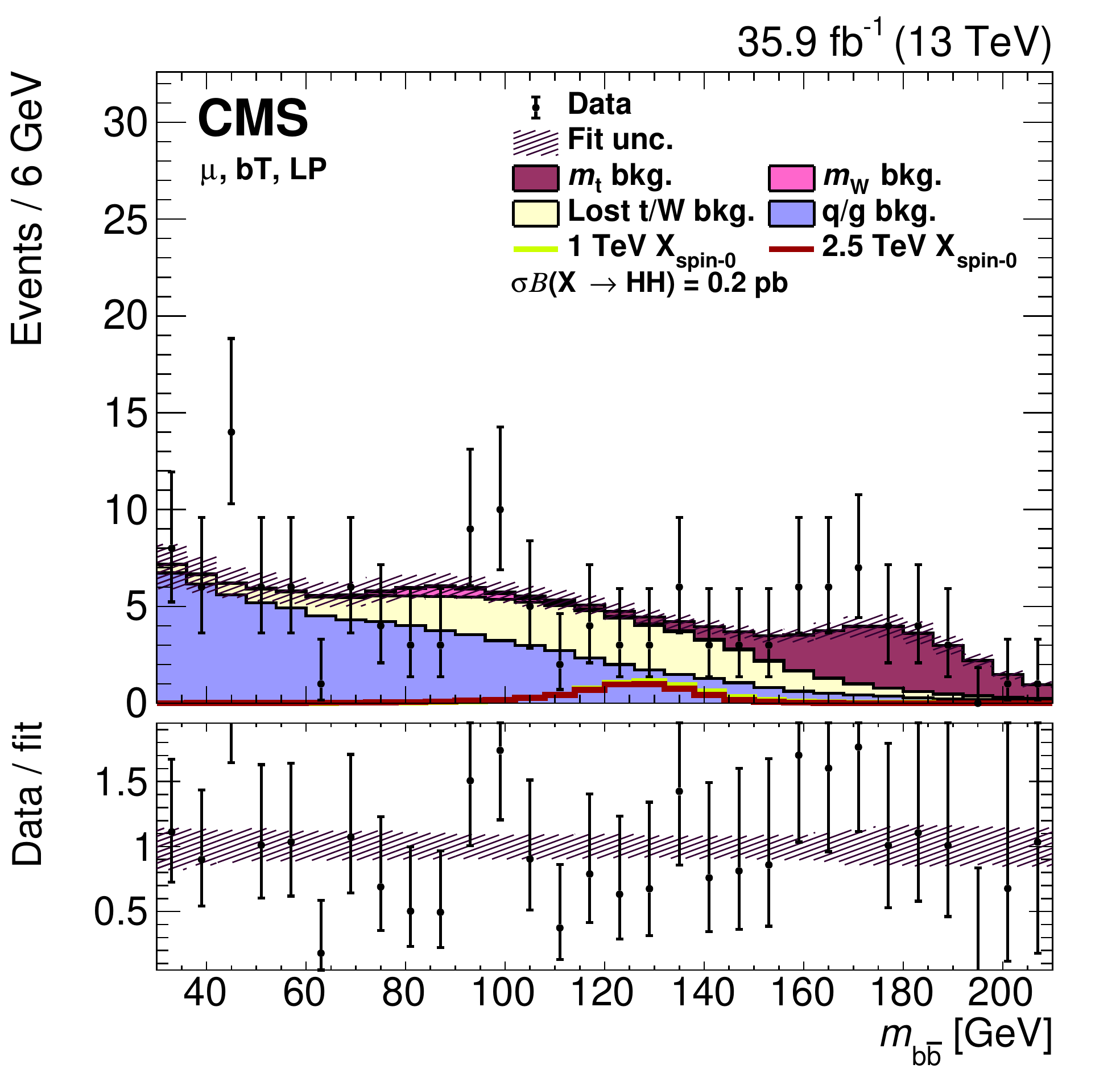}
\includegraphics[width=0.45\textwidth]{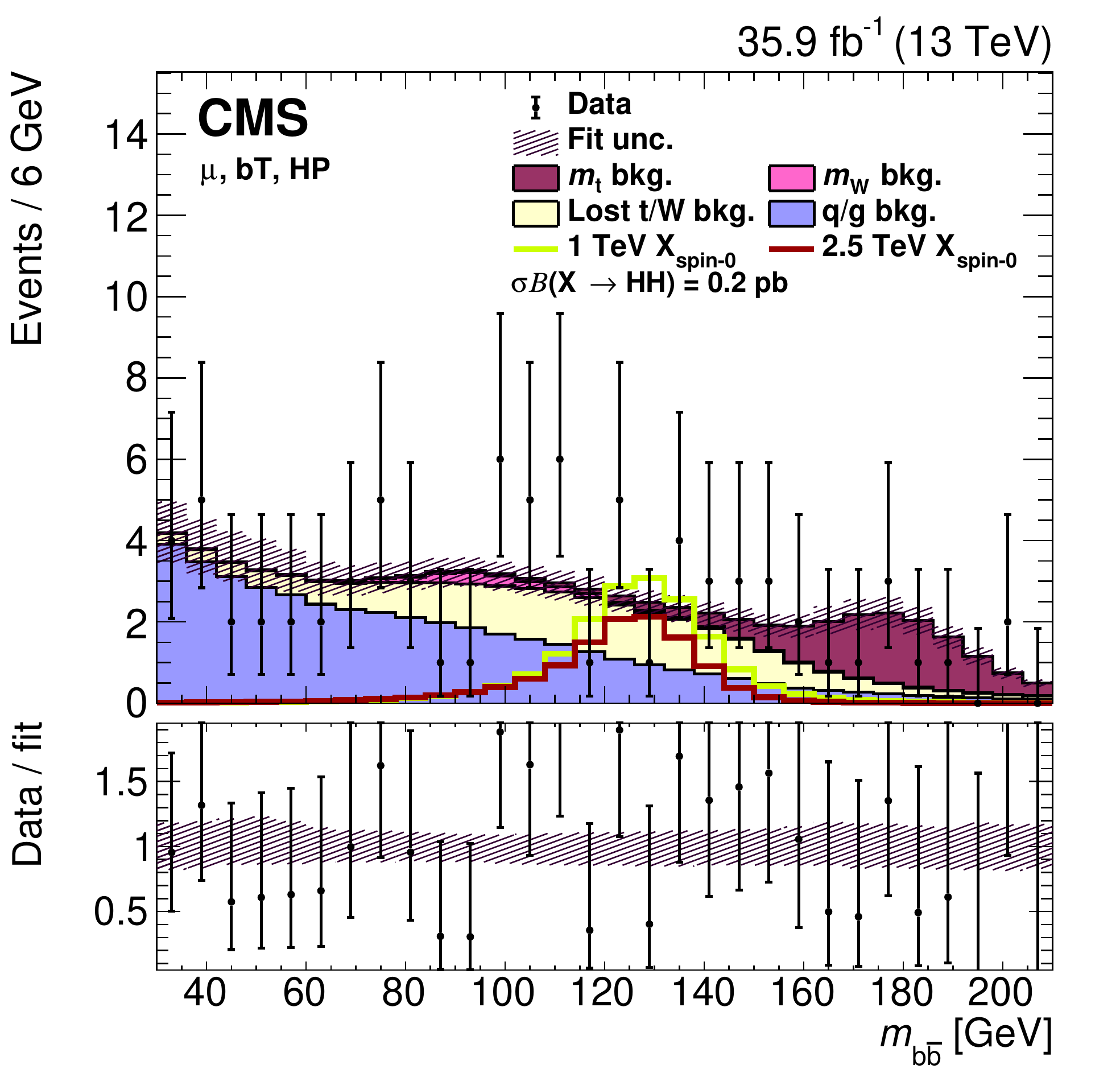}

\caption{The fit result compared to data projected in \mhbb for the muon event categories. The fit result is the filled histogram, with the different colors indicating different background categories. The background shape uncertainty is shown as the hatched band. Example spin-0 signal distributions for $\mx$ of 1 and 2.5\TeV are shown as solid lines, with the product of the cross section and branching fraction to two Higgs bosons set to 0.2\unit{pb}. The lower panels show the ratio of the data to the fit result.}
\label{fig:results_mu_mhbb}
\end{figure}

\begin{figure}[htp!]
\centering
\includegraphics[width=0.45\textwidth]{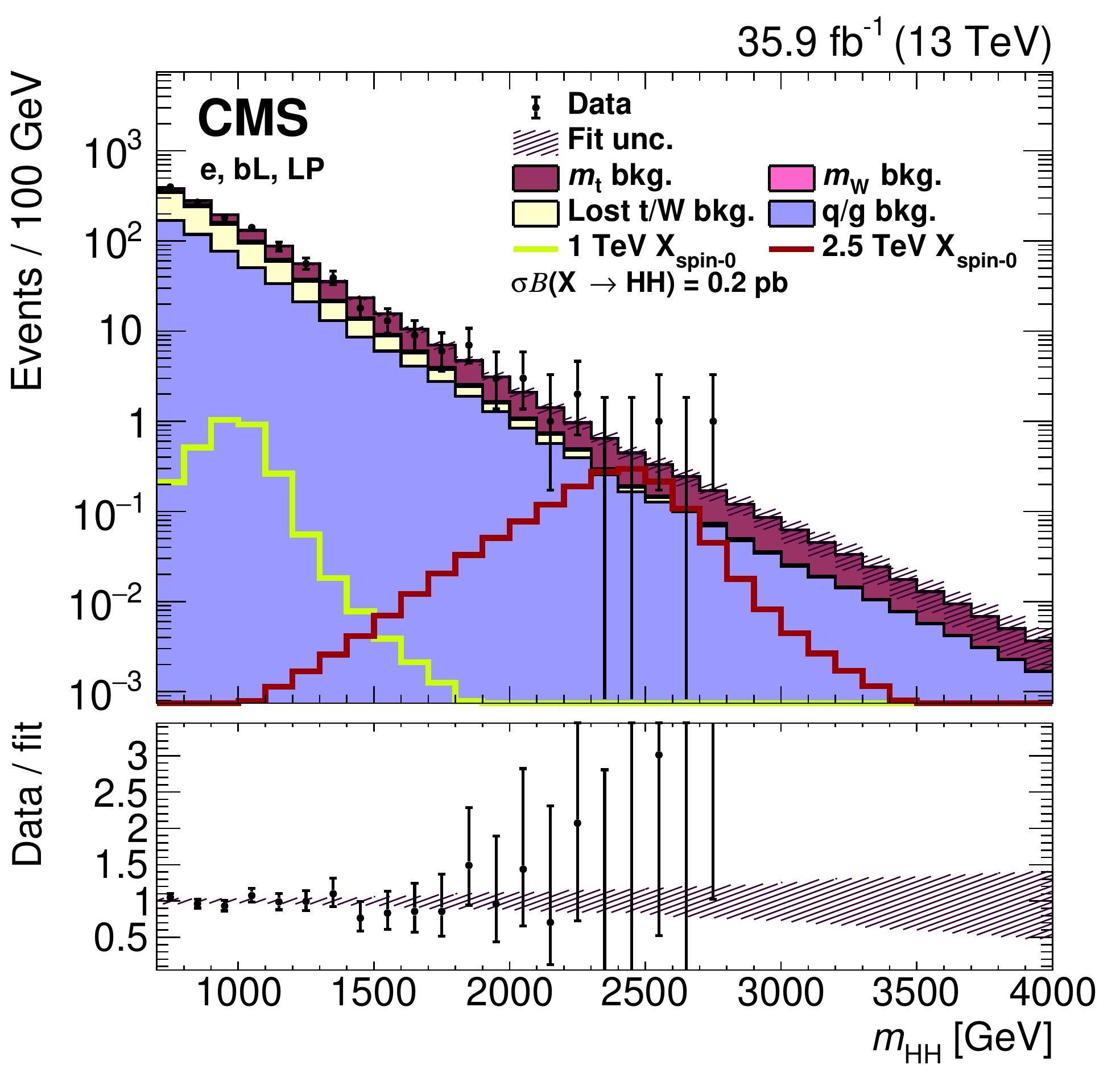}
\includegraphics[width=0.45\textwidth]{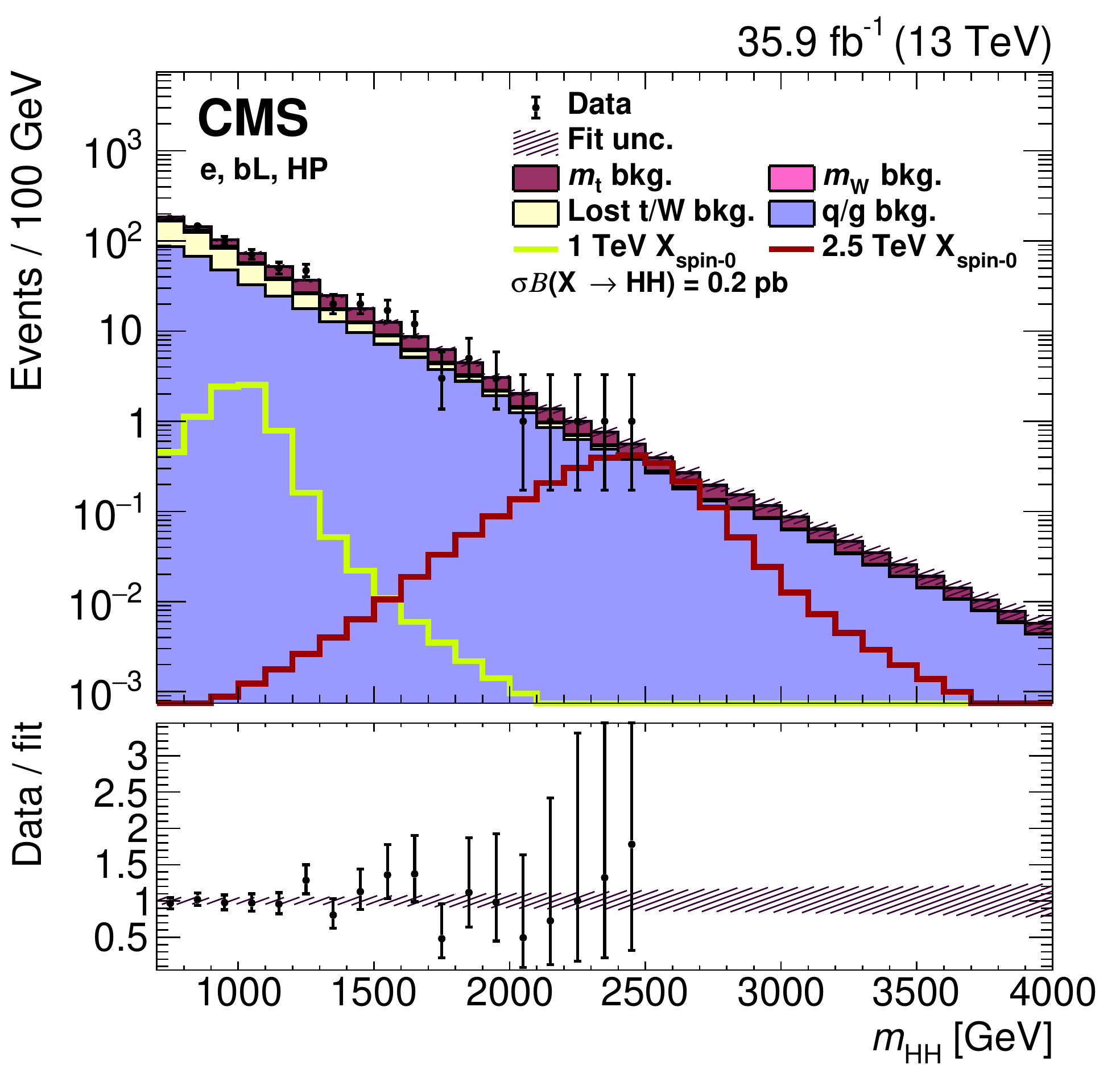}
\includegraphics[width=0.45\textwidth]{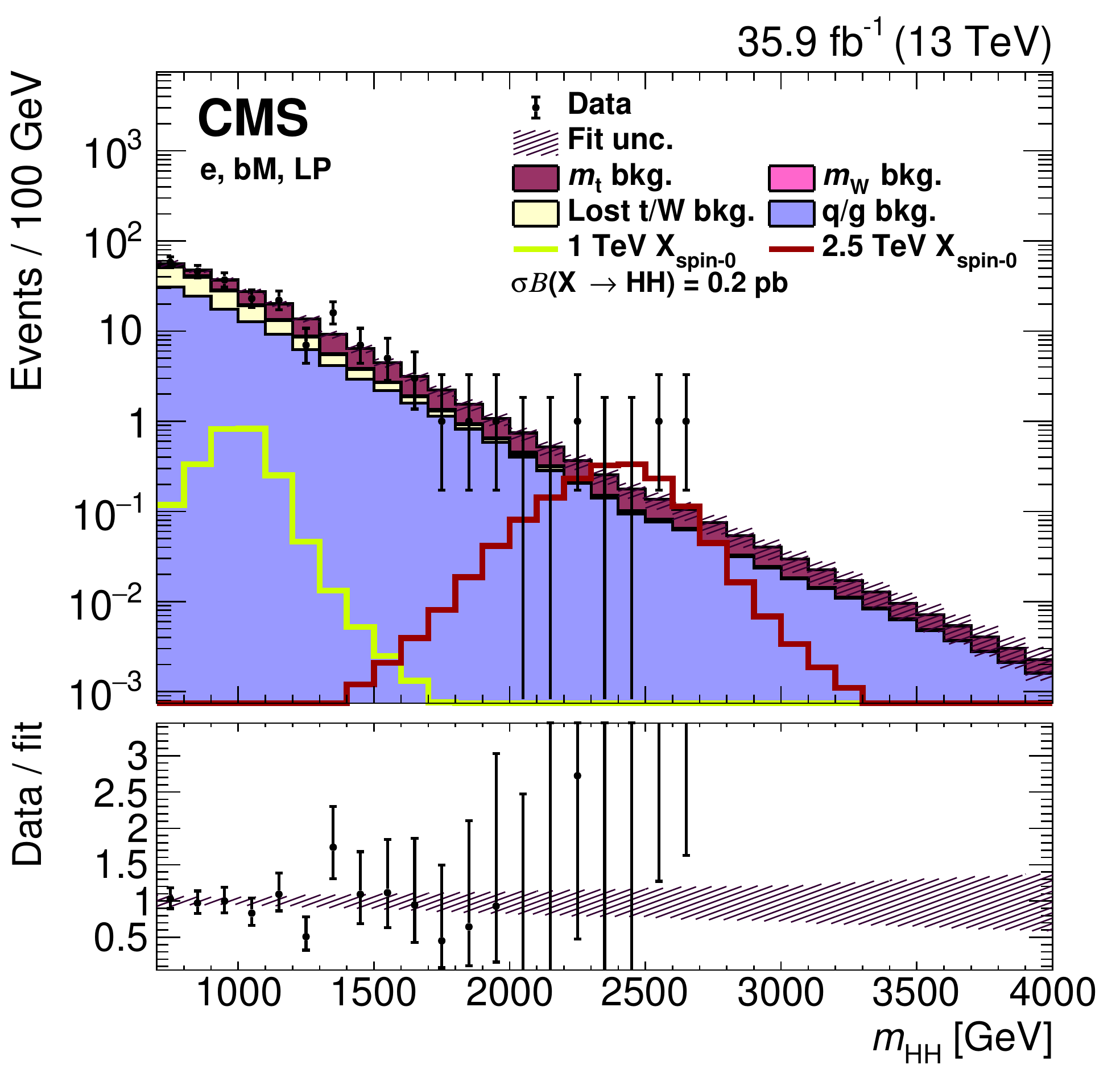}
\includegraphics[width=0.45\textwidth]{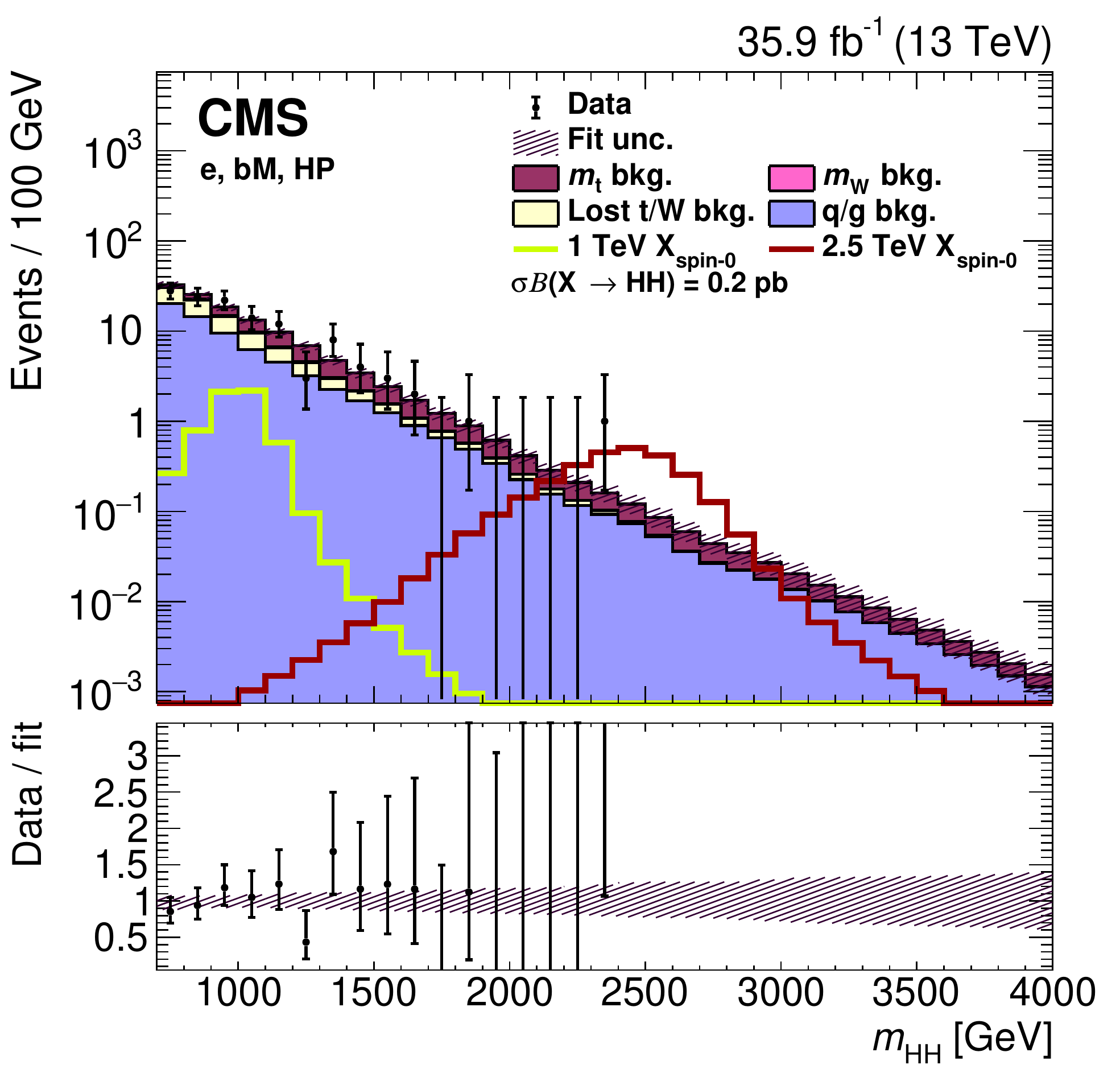}
\includegraphics[width=0.45\textwidth]{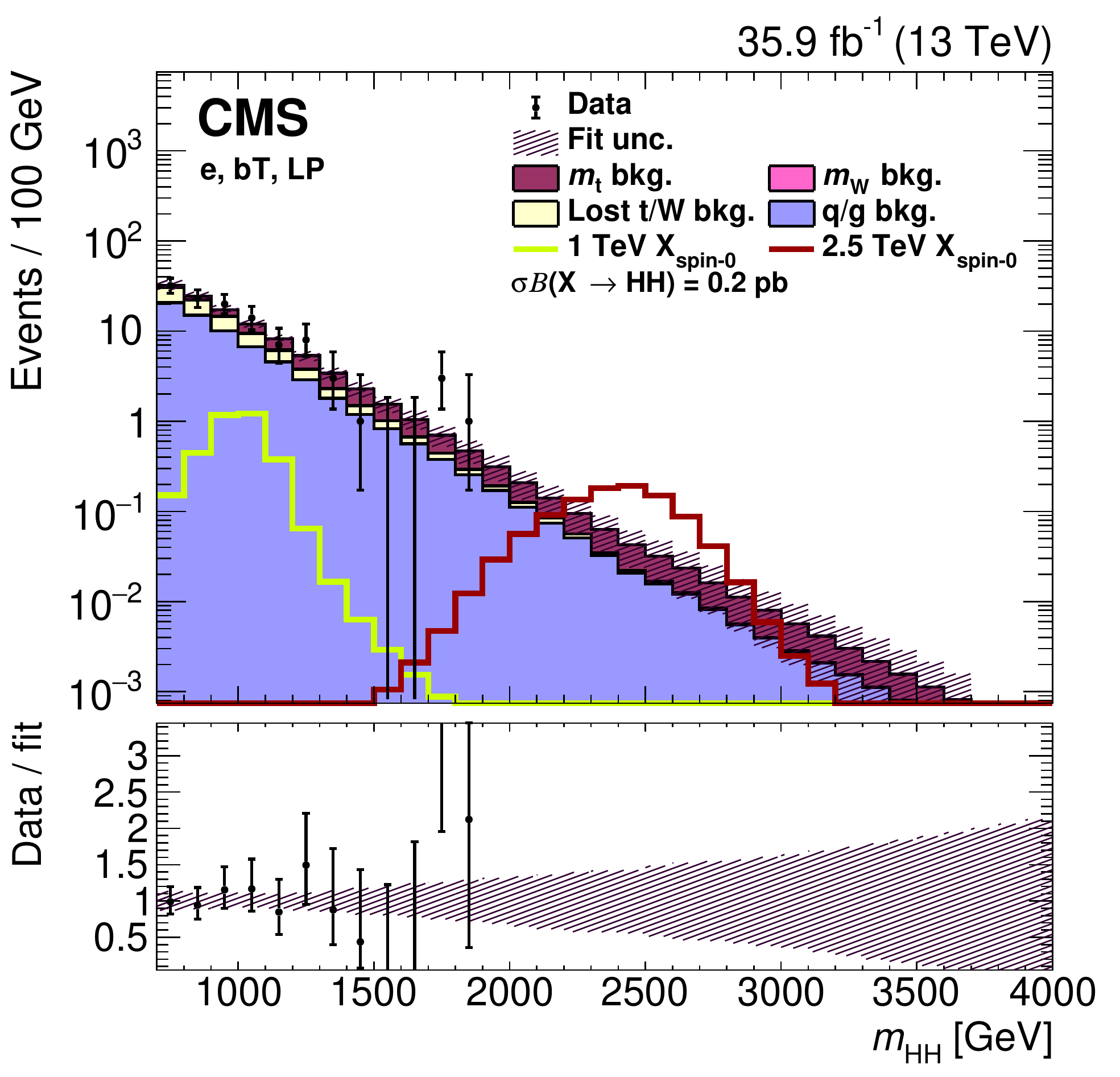}
\includegraphics[width=0.45\textwidth]{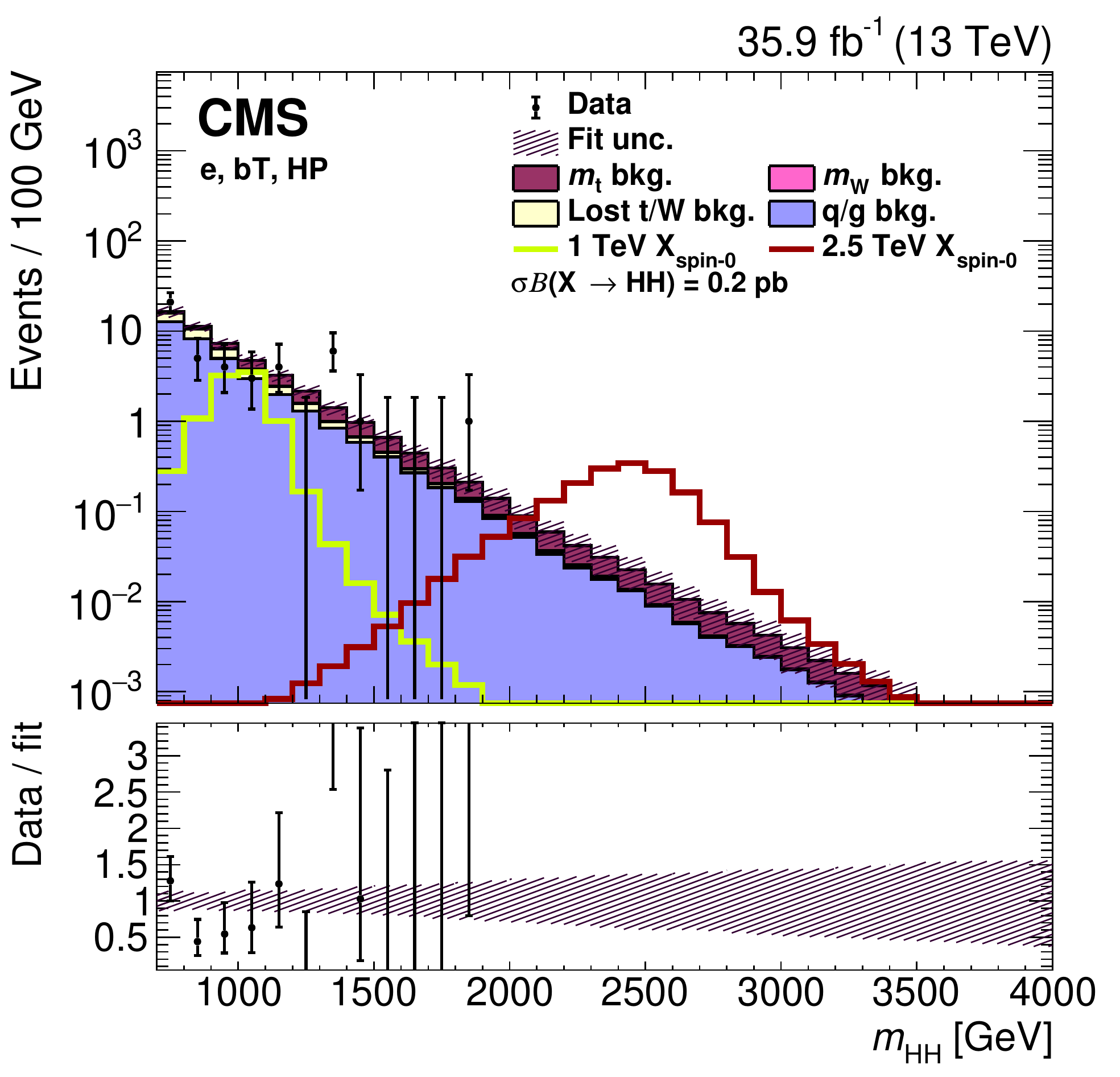}

\caption{The fit result compared to data projected in \mhh for the electron event categories. The fit result is the filled histogram, with the different colors indicating different background categories. The background shape uncertainty is shown as the hatched band. Example spin-0 signal distributions for $\mx$ of 1 and 2.5\TeV are shown as solid lines, with the product of the cross section and branching fraction to two Higgs bosons set to 0.2\unit{pb}. The lower panels show the ratio of the data to the fit result.}
\label{fig:results_e_mhh}
\end{figure}

\begin{figure}[htp!]
\centering
\includegraphics[width=0.45\textwidth]{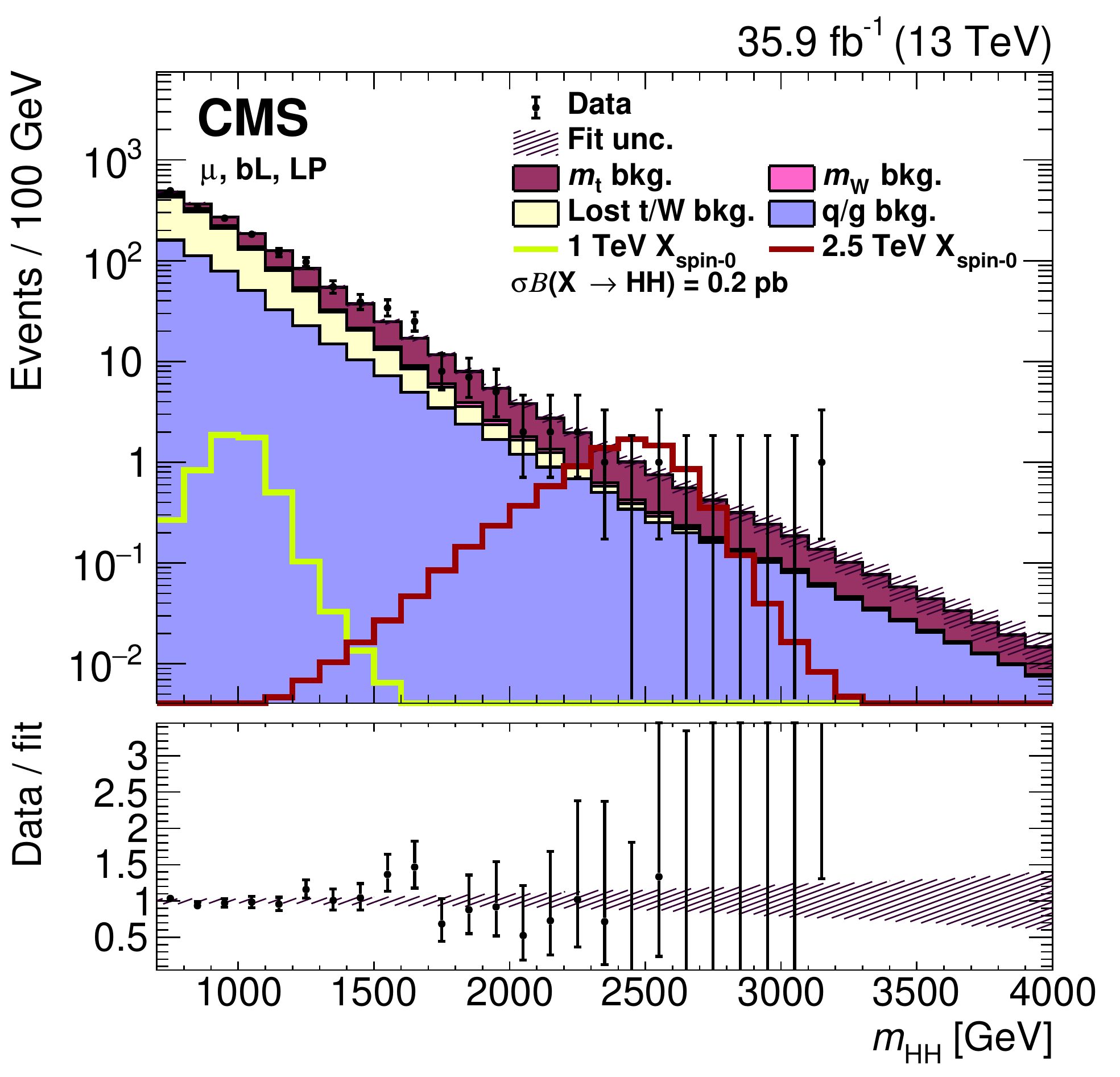}
\includegraphics[width=0.45\textwidth]{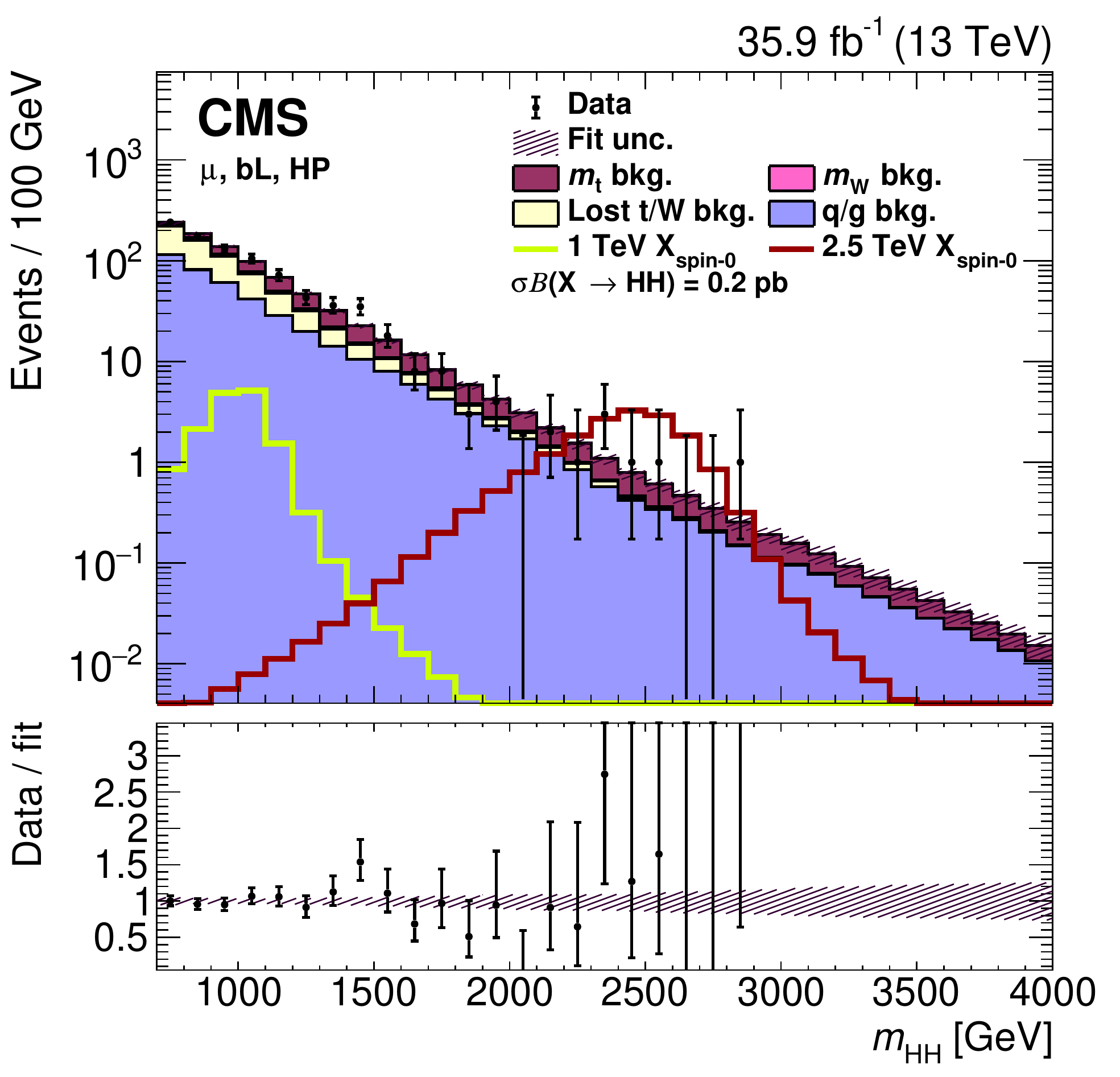}
\includegraphics[width=0.45\textwidth]{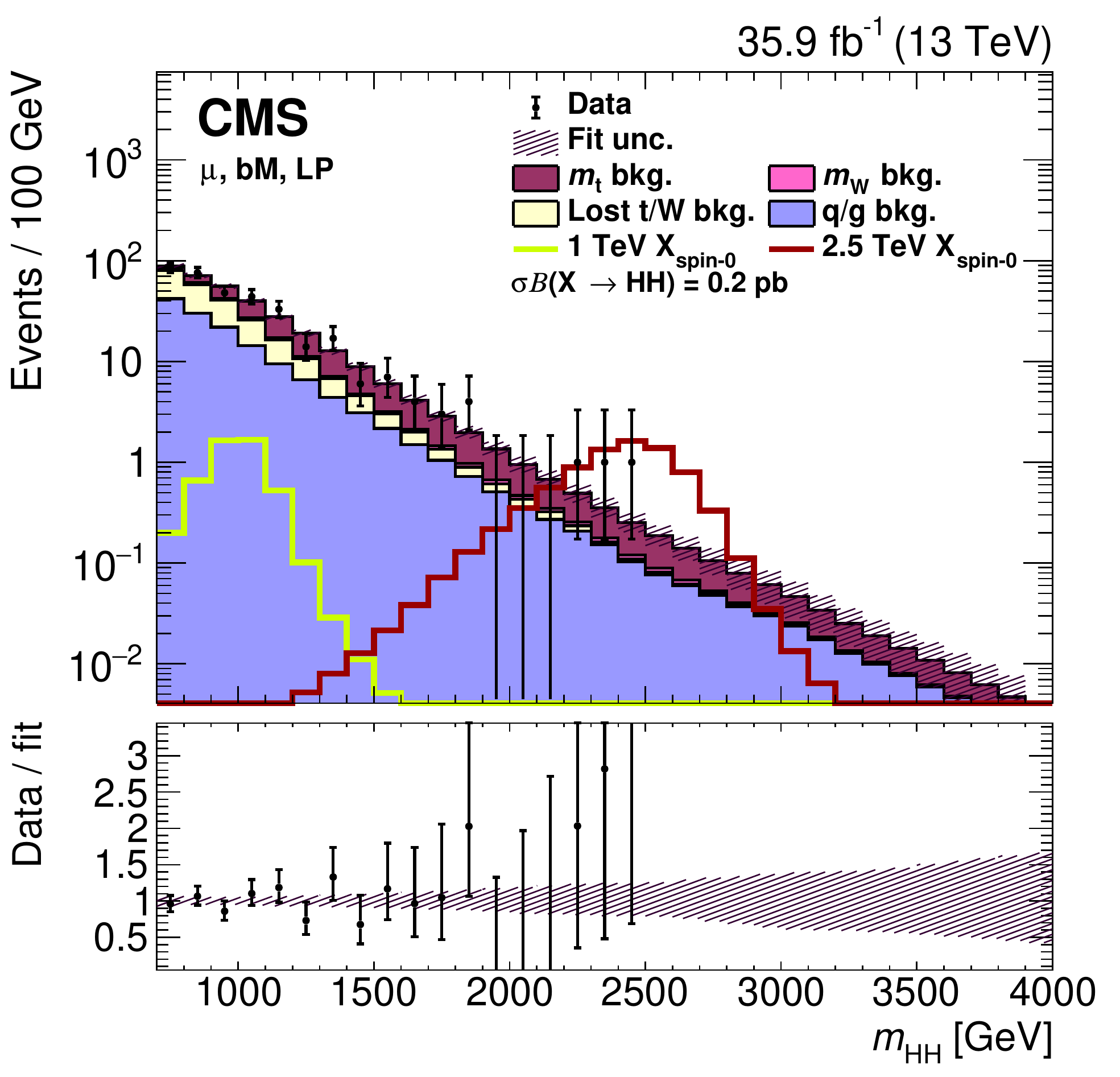}
\includegraphics[width=0.45\textwidth]{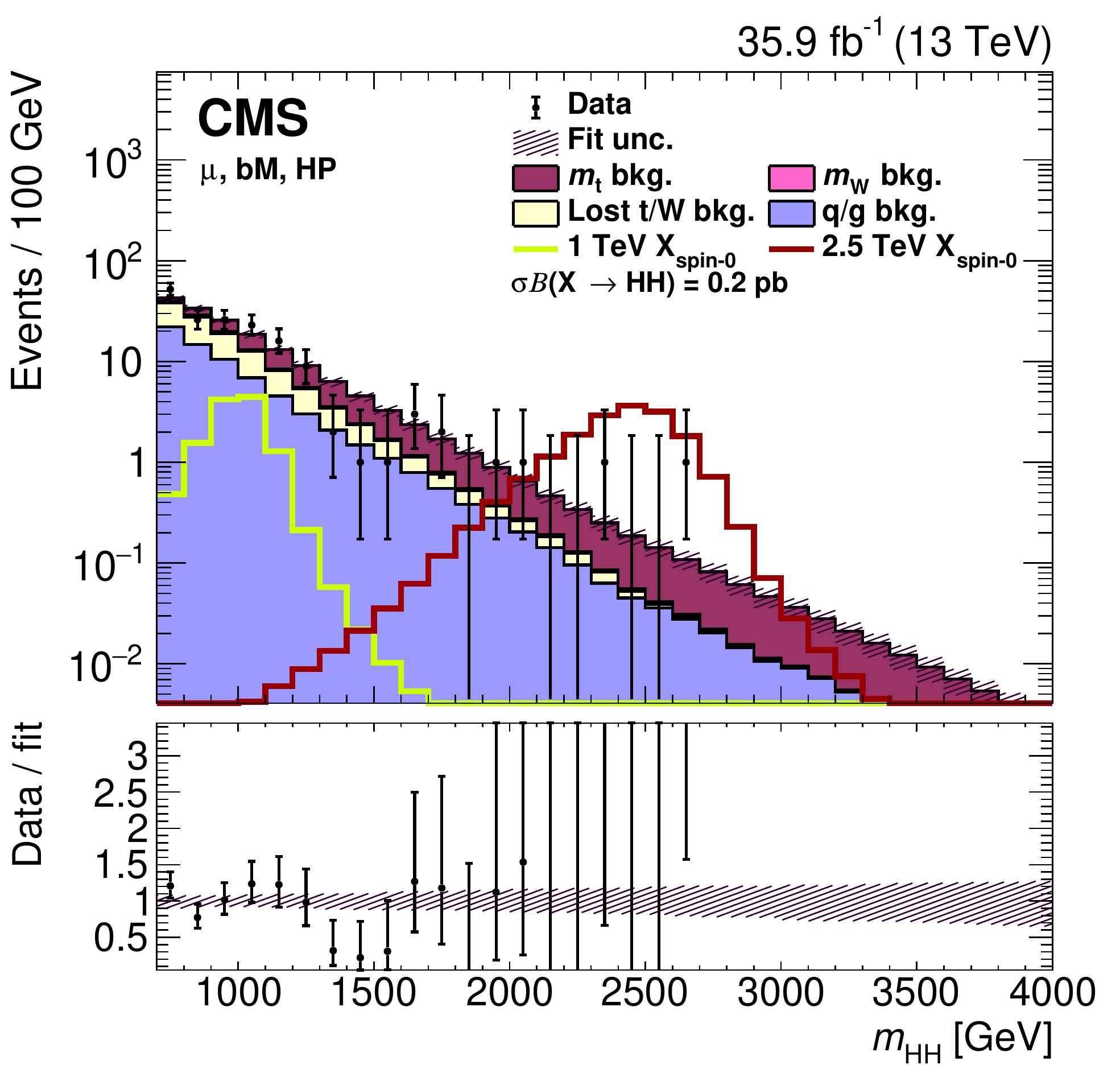}
\includegraphics[width=0.45\textwidth]{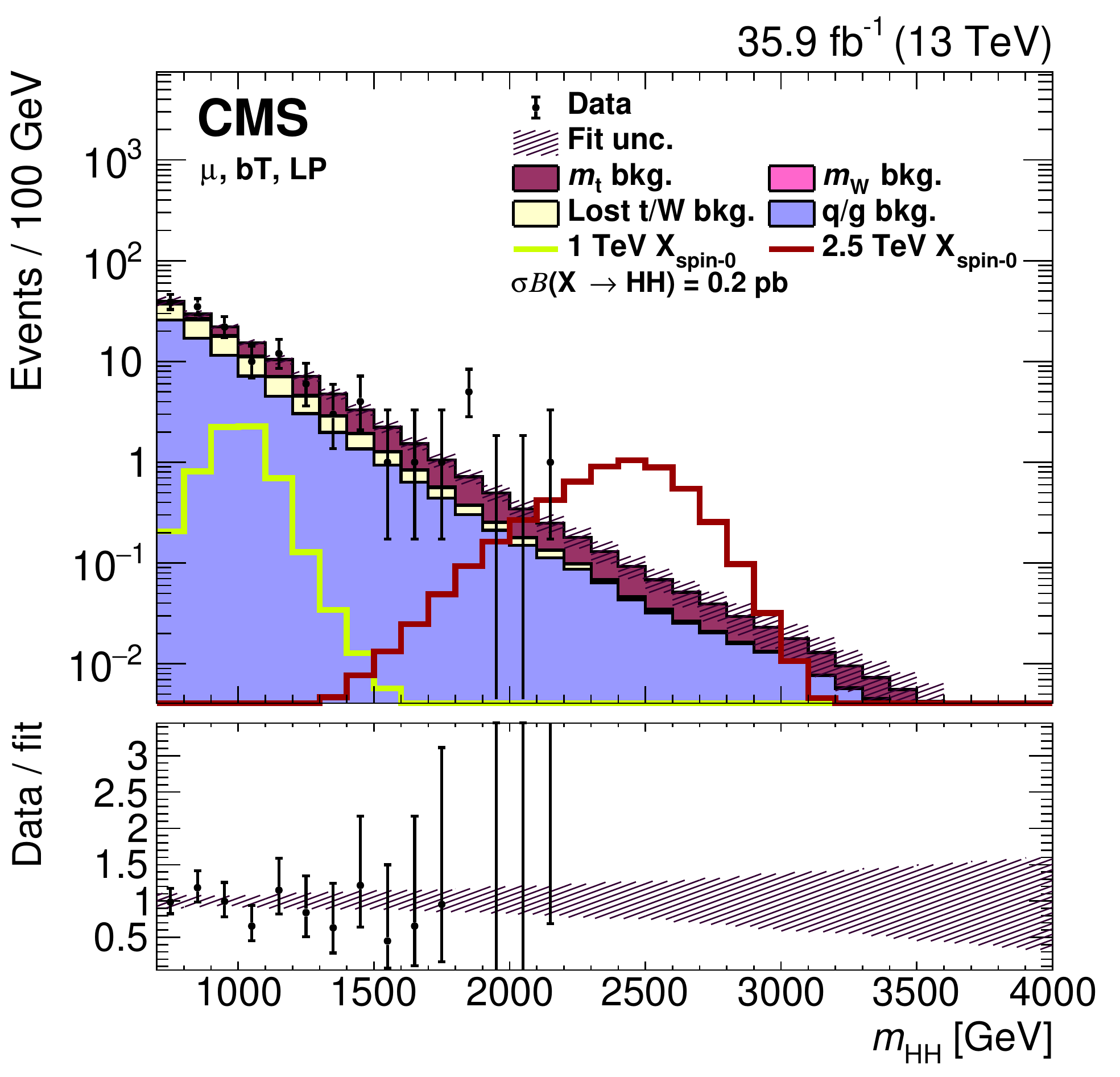}
\includegraphics[width=0.45\textwidth]{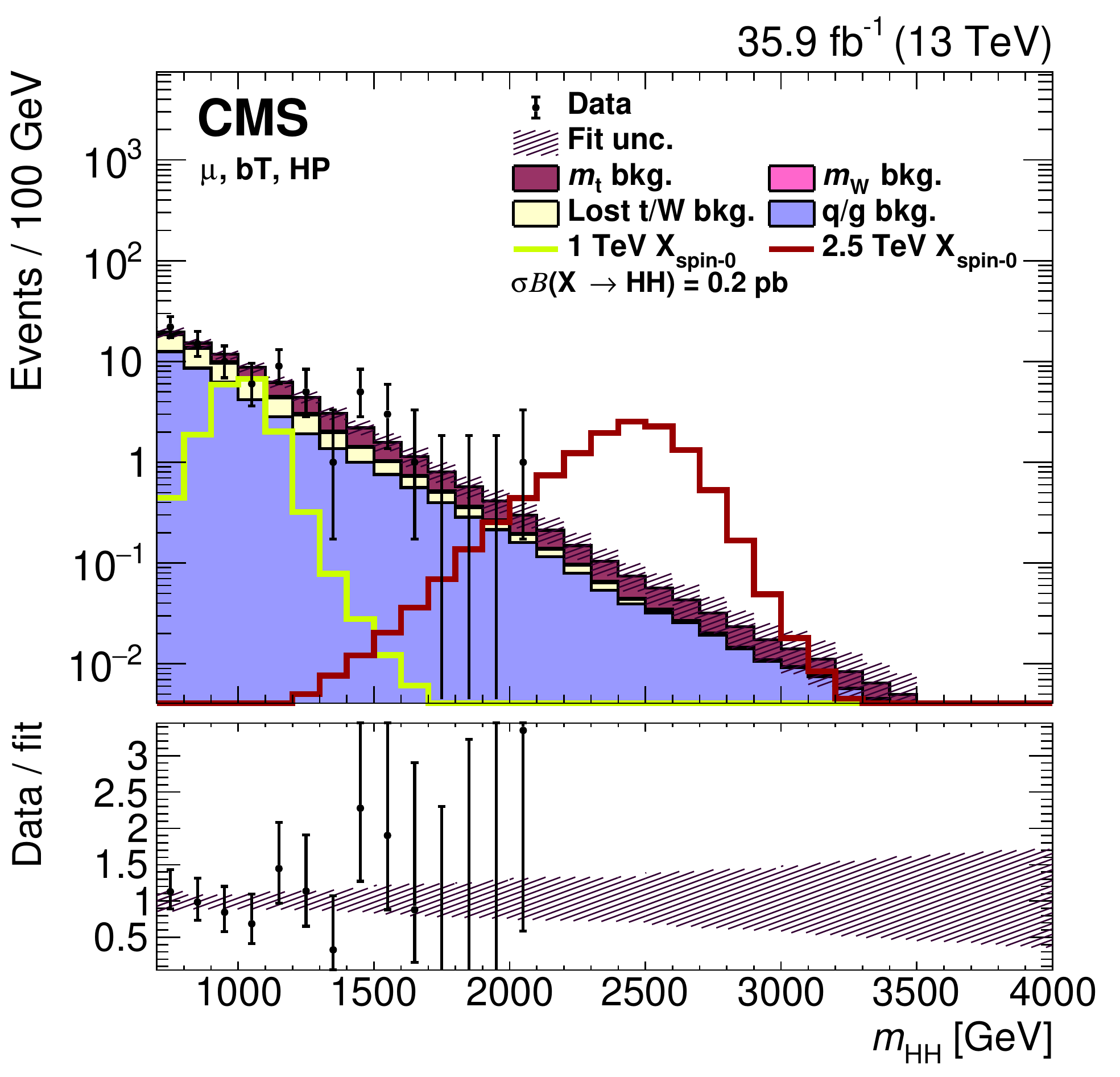}

\caption{The fit result compared to data projected in \mhh for the muon event categories. The fit result is the filled histogram, with the different colors indicating different background categories. The background shape uncertainty is shown as the hatched band. Example spin-0 signal distributions for $\mx$ of 1 and 2.5\TeV are shown as solid lines, with the product of the cross section and branching fraction to two Higgs bosons set to 0.2\unit{pb}. The lower panels show the ratio of the data to the fit result.}
\label{fig:results_mu_mhh}
\end{figure}

The \hbbjet~{\cPqb} tagging efficiency uncertainty is included as a single nuisance parameter that varies the signal normalization in each {\cPqb} tagging category. The uncertainty depends on \mx, with a maximum size of 10, 4, and 4\% for the \bT, \bM, and \bL categories, respectively. The \bL category normalization uncertainty is anticorrelated with the other two uncertainties. A normalization uncertainty is assigned to the efficiency for passing the AK4 jet {\cPqb} tagging veto. The \wqqtau selection efficiency is measured in a \ttbar data sample for $\PW$~bosons decaying to quarks. The uncertainty in this measurement is included as an uncertainty in the HP and LP category relative yields. An additional extrapolation uncertainty is applied because the jets in this sample have lower \pt than those in signal events. The uncertainty depends on \mx, with a maximum value of 7\%  for $\mx=3.5\TeV$. The LP and HP selection efficiency uncertainties are anticorrelated.

\section{Results\label{sec:results}}
The data are interpreted by performing a maximum likelihood fit for a model containing only background processes and one containing both background and signal processes. The background-only fit is found to model the data. We interpret the results as upper limits on the product of the $\PX$ production cross section and the $\PX\to\PH\PH$ branching fraction ($\sigma \mathcal{B}$).

The quality of the fit is quantified with the generalized $\chi^2$ goodness-of-fit test using saturated models~\cite{Baker:1983tu}. The probability distribution function of the test statistic is obtained with pseudo-experiments and the observed value is within the central 68\% quantile of expected results. The best fit values of the nuisance parameters are consistent with the initial uncertainty ranges.

The fit result and the data are projected in \mhbb for each event category in Figs.~\ref{fig:results_e_mhbb} and~\ref{fig:results_mu_mhbb}. The shape is modeled well, with each background category contributing to a specific subset of the mass range. In particular, the resonant peaks associated with \PW~boson and top quark decays are correctly modeled by the fit. Similarly, the projection in \mhh for each event category is shown in Figs.~\ref{fig:results_e_mhh} and~\ref{fig:results_mu_mhh}. Good agreement is found for the entire \mhh mass range.

\begin{figure}[ht]
\centering
\includegraphics[width=0.45\textwidth]{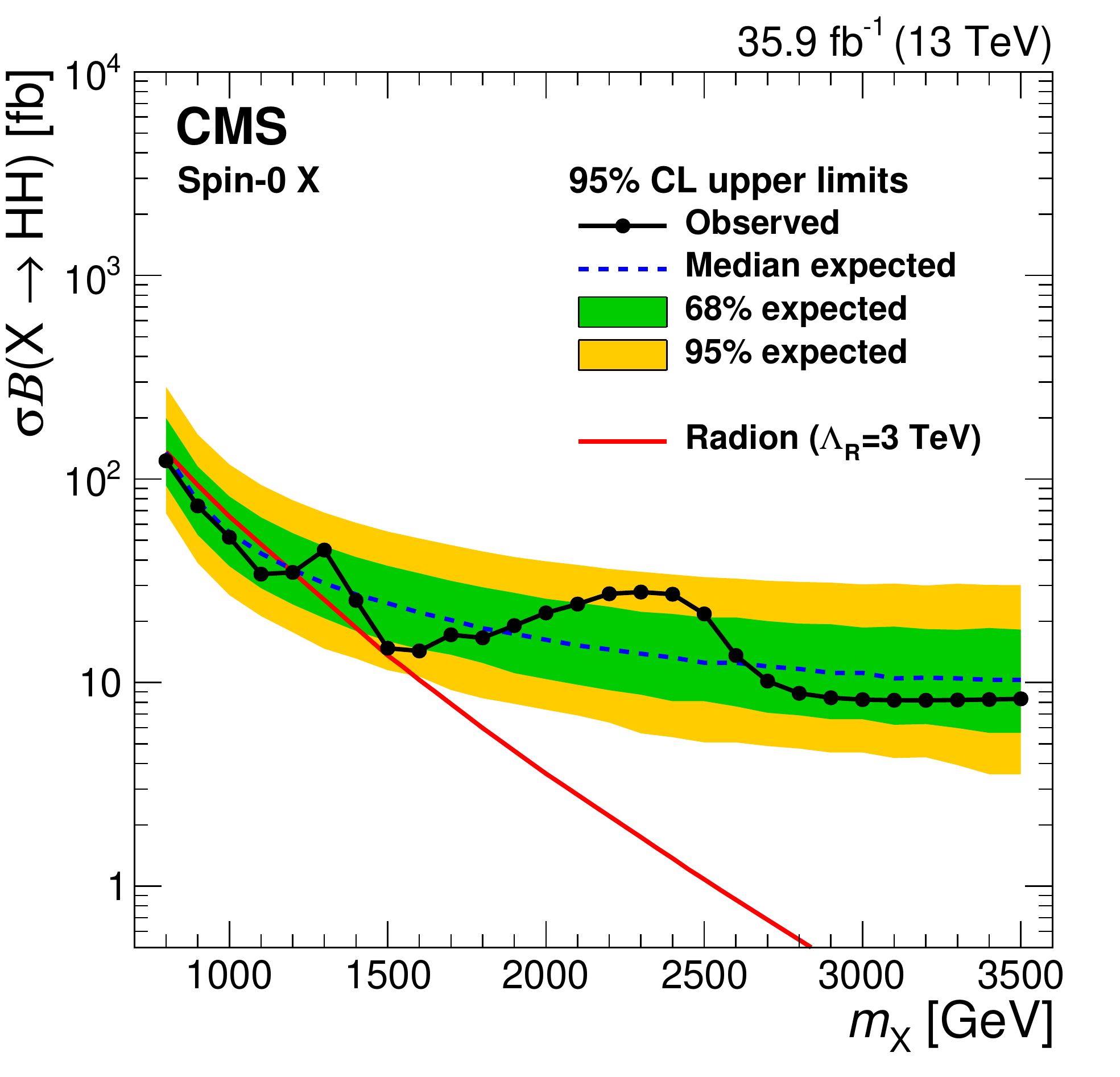}
\includegraphics[width=0.45\textwidth]{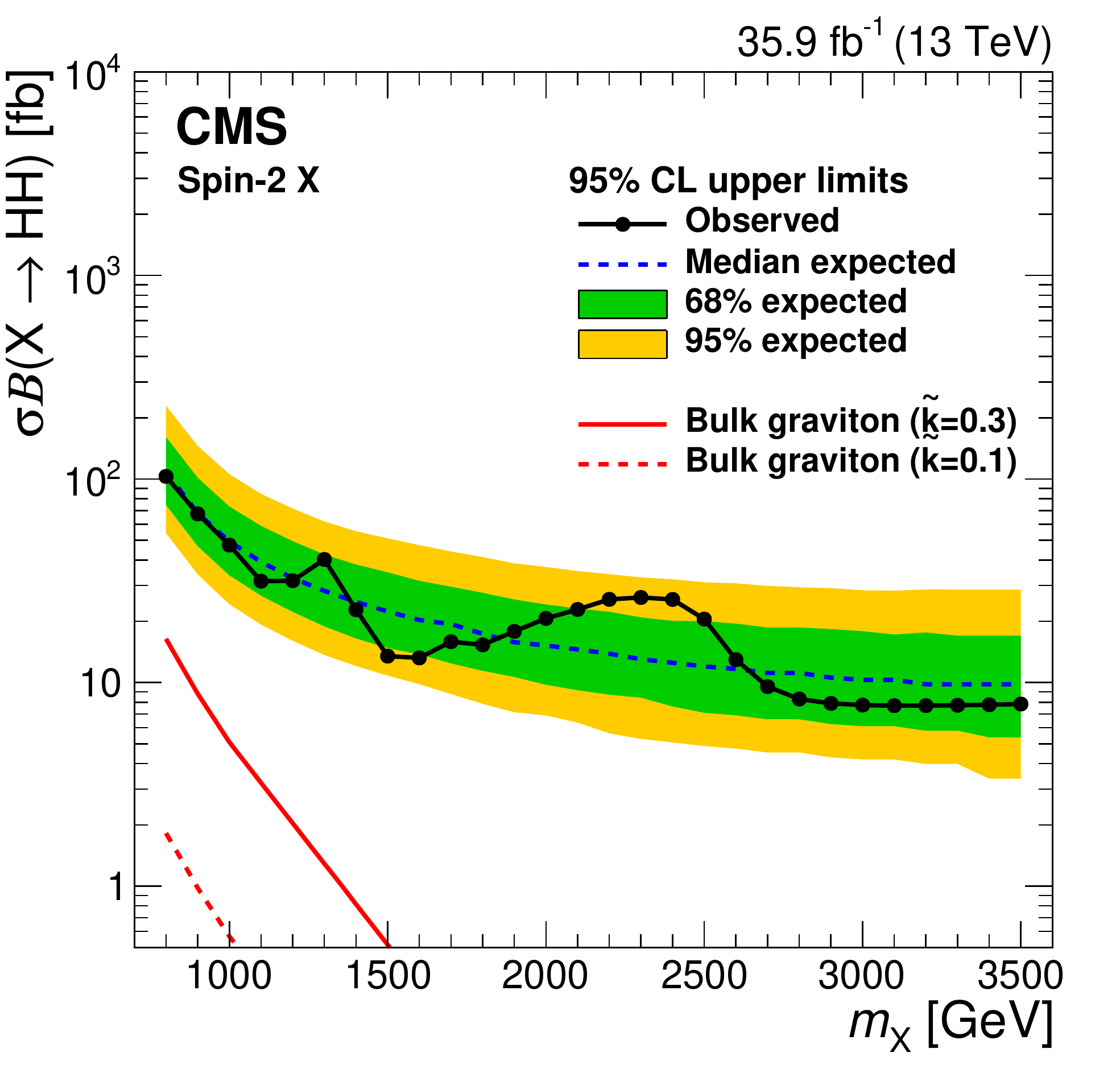}

\caption{Observed and expected 95\% \CL upper limits on the product of the cross section and branching fraction to $\PH\PH$ for a generic spin-0 (left) and spin-2 (right) boson $\PX$, as a function of mass. Example radion and bulk graviton predictions are also shown. The $\PH\PH$ branching fraction is assumed to be 25 and 10\%, respectively. }
\label{fig:results_limits}
\end{figure}

The 95\% confidence level (\CL) upper limits are shown in Fig.~\ref{fig:results_limits} for varying \mx and both the spin-0 and spin-2 boson scenarios. The limits are evaluated using the asymptotic approximation~\cite{Cowan:2010js} of the \CLs method~\cite{Junk:1999kv,Read:2002hq}. The observed exclusion limit is consistent with the expected limit; the most significant deviation between the two is about 1.5 standard deviations at $\mx\approx 2.3\TeV$. The sources of the discrepancy are small excesses in data at high $\mhh$ for the $\mu$, \bM, LP and $\mu$, \bL, HP event categories.
The $\mx = 0.8\TeV$ spin-0 signal is excluded for $\sigma \mathcal{B} > 123\unit{fb}$,  with the exclusion limit strengthening to $\sigma \mathcal{B} > 8.3\unit{fb}$ for $\mx =3.5\TeV$ signal. The higher signal acceptance for spin-2 signal results in stronger constraints on $\sigma \mathcal{B}$: $>$103\unit{fb} for $\mx =0.8\TeV$ signal and $>$7.8\unit{fb} for $\mx =3.5\TeV$ signal.
This search yields the best limits in this decay channel for $\PX\to\PH\PH$ production. It has similar sensitivity to resonances with $\mx\approx1\TeV$ to searches performed in other channels~\cite{Sirunyan:2017isc,Sirunyan:2018fuh}. This search is less sensitive to $\mx\gtrsim1.5\TeV$ resonances because of the degradation of the lepton selection efficiency for events with very large boost.

Predicted radion and bulk graviton cross sections~\cite{Oliveira:2014kla} are also shown in Fig.~\ref{fig:results_limits} in the context of Randall--Sundrum models that allow the SM fields to propagate though the extra dimension. Typical model parameters are chosen as proposed in Ref.~\cite{Gouzevitch:2013qca}. For radions, a branching fraction of 25\% to $\PH\PH$ and an ultraviolet cutoff $\Lambda_R=3\TeV$ are assumed. A 10\% branching fraction is assumed for bulk gravitons, which occurs in scenarios that include significant coupling between the bulk graviton and top quarks. Bulk graviton production cross sections depend on the dimensionless quantity $\widetilde{k}=\sqrt{8\pi}k/M_{\text{Pl}}$, where $k$ is the curvature of the extra dimension and $M_{\text{Pl}}$ is the Planck mass. For this interpretation, we choose $\widetilde{k}=0.1$ and 0.3. For these particular signal parameters the radion and bulk graviton decay widths are larger than the 1\MeV width chosen for signal sample generation, but smaller than the detector resolution.

\section{Summary\label{sec:summary}}
A search has been presented for new particles decaying to a pair of Higgs bosons ($\PH$) where one decays into a bottom quark pair ($\bbbar$) and the other
into two $\PW$~bosons that subsequently decay into a lepton, a neutrino, and a quark pair. The large Lorentz boost of the Higgs bosons leads to the distinct experimental signature of one large-radius jet with substructure consistent with  the decay $\PH\to\bbbar$ and a second large-radius jet with a nearby isolated lepton consistent with the decay $\PH\to\PW\PW^*$. This search uses a sample of proton-proton collisions collected at $\sqrt{s}=13\TeV$ by the CMS detector, corresponding to an integrated luminosity of $35.9\fbinv$. The primary standard model background, top quark pair production, is suppressed by reconstructing the $\PH\PH$ decay chain and applying mass constraints. The signal and background yields are estimated by a two-dimensional template fit in the plane of the \hbbjet mass and the $\PH\PH$ resonance mass. The templates are validated in a variety of data control regions and are shown to model the data well. The data are consistent with the expected standard model background. The results represent upper limits on the product of cross section and branching fraction for new bosons decaying to $\PH\PH$. The observed limit at 95\% confidence level for a spin-0 resonance ranges from  123\unit{fb} at 0.8\TeV to 8.3\unit{fb} at 3.5\TeV, while the limit for a spin-2 resonance is 103\unit{fb} at 0.8\TeV and 7.8\unit{fb} at 3.5\TeV. These are the best results to date from searches for an $\PH\PH$ resonance decaying to this final state. The results are comparable to the results from searches in other channels for resonances with masses below 1.5\TeV.

\begin{acknowledgments}
\hyphenation{Bundes-ministerium Forschungs-gemeinschaft Forschungs-zentren Rachada-pisek} We congratulate our colleagues in the CERN accelerator departments for the excellent performance of the LHC and thank the technical and administrative staffs at CERN and at other CMS institutes for their contributions to the success of the CMS effort. In addition, we gratefully acknowledge the computing centers and personnel of the Worldwide LHC Computing Grid for delivering so effectively the computing infrastructure essential to our analyses. Finally, we acknowledge the enduring support for the construction and operation of the LHC and the CMS detector provided by the following funding agencies: the Austrian Federal Ministry of Education, Science and Research and the Austrian Science Fund; the Belgian Fonds de la Recherche Scientifique, and Fonds voor Wetenschappelijk Onderzoek; the Brazilian Funding Agencies (CNPq, CAPES, FAPERJ, FAPERGS, and FAPESP); the Bulgarian Ministry of Education and Science; CERN; the Chinese Academy of Sciences, Ministry of Science and Technology, and National Natural Science Foundation of China; the Colombian Funding Agency (COLCIENCIAS); the Croatian Ministry of Science, Education and Sport, and the Croatian Science Foundation; the Research Promotion Foundation, Cyprus; the Secretariat for Higher Education, Science, Technology and Innovation, Ecuador; the Ministry of Education and Research, Estonian Research Council via IUT23-4, IUT23-6 and PRG445 and European Regional Development Fund, Estonia; the Academy of Finland, Finnish Ministry of Education and Culture, and Helsinki Institute of Physics; the Institut National de Physique Nucl\'eaire et de Physique des Particules~/~CNRS, and Commissariat \`a l'\'Energie Atomique et aux \'Energies Alternatives~/~CEA, France; the Bundesministerium f\"ur Bildung und Forschung, Deutsche Forschungsgemeinschaft, and Helmholtz-Gemeinschaft Deutscher Forschungszentren, Germany; the General Secretariat for Research and Technology, Greece; the National Research, Development and Innovation Fund, Hungary; the Department of Atomic Energy and the Department of Science and Technology, India; the Institute for Studies in Theoretical Physics and Mathematics, Iran; the Science Foundation, Ireland; the Istituto Nazionale di Fisica Nucleare, Italy; the Ministry of Science, ICT and Future Planning, and National Research Foundation (NRF), Republic of Korea; the Ministry of Education and Science of the Republic of Latvia; the Lithuanian Academy of Sciences; the Ministry of Education, and University of Malaya (Malaysia); the Ministry of Science of Montenegro; the Mexican Funding Agencies (BUAP, CINVESTAV, CONACYT, LNS, SEP, and UASLP-FAI); the Ministry of Business, Innovation and Employment, New Zealand; the Pakistan Atomic Energy Commission; the Ministry of Science and Higher Education and the National Science Centre, Poland; the Funda\c{c}\~ao para a Ci\^encia e a Tecnologia, Portugal; JINR, Dubna; the Ministry of Education and Science of the Russian Federation, the Federal Agency of Atomic Energy of the Russian Federation, Russian Academy of Sciences, the Russian Foundation for Basic Research, and the National Research Center ``Kurchatov Institute"; the Ministry of Education, Science and Technological Development of Serbia; the Secretar\'{\i}a de Estado de Investigaci\'on, Desarrollo e Innovaci\'on, Programa Consolider-Ingenio 2010, Plan Estatal de Investigaci\'on Cient\'{\i}fica y T\'ecnica y de Innovaci\'on 2013--2016, Plan de Ciencia, Tecnolog\'{i}a e Innovaci\'on 2013--2017 del Principado de Asturias, and Fondo Europeo de Desarrollo Regional, Spain; the Ministry of Science, Technology and Research, Sri Lanka; the Swiss Funding Agencies (ETH Board, ETH Zurich, PSI, SNF, UniZH, Canton Zurich, and SER); the Ministry of Science and Technology, Taipei; the Thailand Center of Excellence in Physics, the Institute for the Promotion of Teaching Science and Technology of Thailand, Special Task Force for Activating Research and the National Science and Technology Development Agency of Thailand; the Scientific and Technical Research Council of Turkey, and Turkish Atomic Energy Authority; the National Academy of Sciences of Ukraine, and State Fund for Fundamental Researches, Ukraine; the Science and Technology Facilities Council, UK; the US Department of Energy, and the US National Science Foundation.
Individuals have received support from the Marie-Curie program and the European Research Council and Horizon 2020 Grant, contract Nos.\ 675440 and 765710 (European Union); the Leventis Foundation; the A.P.\ Sloan Foundation; the Alexander von Humboldt Foundation; the Belgian Federal Science Policy Office; the Fonds pour la Formation \`a la Recherche dans l'Industrie et dans l'Agriculture (FRIA-Belgium); the Agentschap voor Innovatie door Wetenschap en Technologie (IWT-Belgium); the F.R.S.-FNRS and FWO (Belgium) under the ``Excellence of Science -- EOS" -- be.h project n.\ 30820817; the Beijing Municipal Science \& Technology Commission, No. Z181100004218003; the Ministry of Education, Youth and Sports (MEYS) of the Czech Republic; the Lend\"ulet (``Momentum") Programme and the J\'anos Bolyai Research Scholarship of the Hungarian Academy of Sciences, the New National Excellence Program \'UNKP, the NKFIA research grants 123842, 123959, 124845, 124850, 125105, 128713, 128786, and 129058 (Hungary); the Council of Scientific and Industrial Research, India; the HOMING PLUS program of the Foundation for Polish Science, cofinanced from European Union, Regional Development Fund, the Mobility Plus program of the Ministry of Science and Higher Education, the National Science Center (Poland), contracts Harmonia 2014/14/M/ST2/00428, Opus 2014/13/B/ST2/02543, 2014/15/B/ST2/03998, and 2015/19/B/ST2/02861, Sonata-bis 2012/07/E/ST2/01406; the National Priorities Research Program by Qatar National Research Fund; the Programa de Excelencia Mar\'{i}a de Maeztu, and the Programa Severo Ochoa del Principado de Asturias; the Thalis and Aristeia programs cofinanced by EU-ESF, and the Greek NSRF; the Rachadapisek Sompot Fund for Postdoctoral Fellowship, Chulalongkorn University, and the Chulalongkorn Academic into Its 2nd Century Project Advancement Project (Thailand); the Welch Foundation, contract C-1845; and the Weston Havens Foundation (USA).
\end{acknowledgments}

\bibliography{auto_generated}
\cleardoublepage \appendix\section{The CMS Collaboration \label{app:collab}}\begin{sloppypar}\hyphenpenalty=5000\widowpenalty=500\clubpenalty=5000\input{B2G-18-008-authorlist.tex}\end{sloppypar}
\end{document}

%% file: B2G-18-008-authorlist.tex
\vskip\cmsinstskip
\textbf{Yerevan Physics Institute, Yerevan, Armenia}\\*[0pt]
A.M.~Sirunyan$^{\textrm{\dag}}$, A.~Tumasyan
\vskip\cmsinstskip
\textbf{Institut f\"{u}r Hochenergiephysik, Wien, Austria}\\*[0pt]
W.~Adam, F.~Ambrogi, T.~Bergauer, J.~Brandstetter, M.~Dragicevic, J.~Er\"{o}, A.~Escalante~Del~Valle, M.~Flechl, R.~Fr\"{u}hwirth\cmsAuthorMark{1}, M.~Jeitler\cmsAuthorMark{1}, N.~Krammer, I.~Kr\"{a}tschmer, D.~Liko, T.~Madlener, I.~Mikulec, N.~Rad, J.~Schieck\cmsAuthorMark{1}, R.~Sch\"{o}fbeck, M.~Spanring, D.~Spitzbart, W.~Waltenberger, J.~Wittmann, C.-E.~Wulz\cmsAuthorMark{1}, M.~Zarucki
% ...
\vskip\cmsinstskip
\dag: Deceased\\
1:  Also at Vienna University of Technology, Vienna, Austria\\
% ...
